\documentclass[sigconf]{acmart}

\usepackage{graphicx}
\usepackage{amsmath}
\usepackage{latexsym}
\usepackage{subfigure}
\usepackage{csquotes}
\usepackage{fancyhdr}
\usepackage{epstopdf}
\usepackage{url}
\usepackage{array,color}
\usepackage{algorithm}
\usepackage{algpseudocode}
\usepackage{epsfig}
\usepackage{epstopdf}
\usepackage{amssymb}
\usepackage{amsthm}
\usepackage{multirow} 
\usepackage{blindtext}
\usepackage{caption} 

\algnewcommand\algorithmicswitch{\textbf{switch}}
\algnewcommand\algorithmiccase{\textbf{Case}}
\algnewcommand\algorithmicassert{\texttt{assert}}
\algnewcommand\Assert[1]{\State \algorithmicassert(#1)}

\algdef{SE}[SWITCH]{Switch}{EndSwitch}[1]{\algorithmicswitch\ #1\ \algorithmicdo}{\algorithmicend\ \algorithmicswitch}%
\algdef{SE}[CASE]{Case}{EndCase}[1]{\algorithmiccase\ #1}{\algorithmicend\ \algorithmiccase}%
\algtext*{EndSwitch}%
\algtext*{EndCase}%

\newcommand{\topoa}{ATT}

\newcommand{\solution}{RetroFlow}
\newtheorem{mythe}{Theorem}

\newcommand{\qedd}{\hfill $\square$}
\begin{document}
\copyrightyear{2019} 
\acmYear{2019} 
\setcopyright{acmcopyright}
\acmConference[IWQoS '19]{IEEE/ACM International Symposium on Quality of Service}{June 24--25, 2019}{Phoenix, AZ, USA}
\acmBooktitle{IEEE/ACM International Symposium on Quality of Service (IWQoS '19), June 24--25, 2019, Phoenix, AZ, USA}
\acmPrice{15.00}
\acmDOI{10.1145/3326285.3329036}
\acmISBN{978-1-4503-6778-3/19/06}

\title[\solution: Maintaining Control Resiliency and Flow Programmab\\ -ility for Software-Defined WANs]{\solution: Maintaining Control Resiliency and Flow Programmability for Software-Defined WANs}

\author{Zehua Guo}
\authornote{Corresponding author: Zehua Guo (guolizihao@hotmail.com) affiliated with Beijing Institute of Technology. This work was done when the first author was a postdoctoral research associate and the second and third authors were visiting PhD students at the University of Minnesota.}
\affiliation{%
  \institution{University of Minnesota Twin Cities}
}
\affiliation{%
\institution{Beijing Institute of Technology}
}

\author{Wendi Feng}
\affiliation{%
  \institution{Beijing University of Posts and Telecommunications}
}

\author{Sen Liu}
\affiliation{%
  \institution{Central South University}
}

\author{Wenchao Jiang}
\affiliation{%
  \institution{University of Minnesota Twin Cities}
}

\author{Yang  Xu}
\affiliation{%
  \institution{Fudan University}
}

\author{Zhi-Li Zhang}
\affiliation{%
  \institution{University of Minnesota Twin Cities}
}

\renewcommand{\shortauthors}{Zehua Guo, \textit{et al.}}


%

\begin{abstract}
Providing resilient network control is a critical concern for deploying Software-Defined Networking (SDN) into Wide-Area Networks (WANs). For performance reasons, a Software-Defined WAN is divided into multiple domains controlled by multiple controllers with a logically centralized view. Under controller failures, we need to remap the control of offline switches from failed controllers to other active controllers. Existing solutions could either overload active controllers to interrupt their normal operations or degrade network performance because of increasing the controller-switch communication overhead. In this paper, we propose \solution \ to achieve low communication overhead without interrupting the normal processing of active controllers during controller failures. By intelligently configuring a set of selected offline switches working under the legacy routing mode, \solution \ relieves the active controllers from controlling the selected offline switches while maintaining the flow programmability (e.g., the ability to change paths of flows) of SDN. \solution \ also smartly transfers the control of offline switches with the SDN routing mode to active controllers to minimize the communication overhead from these offline switches to the active controllers. Simulation results show that compared with the baseline algorithm, \solution \ can reduce the communication overhead up to 52.6\% during a moderate controller failure by recovering 100\% flows from offline switches and can reduce the communication overhead up to 61.2\% during a serious controller failure by setting to recover 90\% of flows from offline switches.
\end{abstract}

\ccsdesc[500]{Networks~Programmable networks}
\ccsdesc[500]{Computer systems organization~Maintainability and maintenance}
\ccsdesc[500]{Networks~Wide area networks}

\keywords{software-defined networking, control plane, resiliency, programmability, wide area networks, hybrid routing}

\maketitle
\setlength\abovedisplayskip{0pt}
\setlength\belowdisplayskip{0pt}



\section{Introduction}
Software-Defined Networking (SDN) has been deployed in real networks \cite{uber}\cite{jain2013b4}. One critical scenario for the SDN is Wide-Area Networks (WANs), known as the SD-WANs. For instance, as one of the world's largest Internet service provider, AT\&T has softwarized 65\% of its WAN with programmable devices (e.g., SDN switches) by 2018 and plans to improve the ratio to 75\% by 2020 \cite{att}. In the future, most of the network infrastructure in WANs (e.g., switches and routers) will be replaced by programmable devices.

In SD-WANs, the network is usually divided into multiple domains to achieve low latency control given the large scale of a WAN and the huge number of SDN switches in it \cite{hu2018multi}. Each domain usually has an SDN controller that can quickly reply to requests from all the SDN switches within the domain. The controllers from different domains are physically distributed, but they can achieve a logically centralized control by the synchronization among them to maintain a consistent network view \cite{guo2014improving}.  



Control resiliency is a critical concern for SD-WANs. Essentially, an SDN controller is a network software installed in a physical server or a virtual machine. Due to some unexpected issues (e.g., hardware/software bugs, power failure), one controller could accidentally fail, and then all of its connected switches are out of control, which we refer to as the \emph{offline switches}. Existing solutions to maintain the control resiliency of SDN can be categorized into two classes: (1) controller placement and (2) switch remapping. Solutions in the former category carefully choose physical locations of controllers to optimize the control performance under controller failures, such as minimizing the latency between backup controllers and switches \cite{hock2013pareto_2}\cite{tanha2016enduring} and/or minimizing the latency among the main controller, backup controllers, and switches \cite{killi2017capacitated}. These solutions are usually based on some unrealistic assumptions, such as the same control cost and unlimited capability of controllers, which are far from practical. In contrast, the switch remapping approaches propose to dynamically shift the control of offline switches to other active controllers \cite{hu2018adaptive}. However, in an almost saturated SDN, controllers almost reach their processing limits. There will be little room left in active controllers to accept offline switches from the failed controllers without overloading the active controllers, which otherwise can degrade the performance (e.g., increasing the communication overhead) or even cause the cascading controller failure \cite{hu2018adaptive}\cite{yao2013cascading}\cite{hu2019dynamic}.


In this paper, we present a feature available in existing commercial SDN switches that shed light on this issue. Existing commercial SDN switches (e.g., Brocade MLX-8 PE \cite{PE}) can freely change between two routing modes, the SDN mode and the legacy mode. The former relies on the SDN controller's decision to process flows while the latter processes flows using its traditional routing table without consulting the controller. Inspired by the feature in commercial devices, we propose to configure switches in hybrid modes so that we can enjoy the flow programmability (e.g., the ability to change the paths of flows) brought by the SDN mode while avoiding the out-of-control disasters coming with the offline switches during the controller failures.

To get the best of both worlds, we have overcome two main challenges. First, configuring SDN switches to work in the legacy mode will decrease the flow programmability of SDN. We carefully choose a set of offline switches working in the legacy mode to maximize the number of programmable flows that have at least one alternative path to forward. Second, the switch-controller mapping affects the communication overhead from offline switches to active controllers. For the remaining offline switches that still work in the SDN mode, we carefully remap them to active controllers, which is a complex optimization problem restricted to the switches' control cost (e.g., the per-flow state pulling \cite{van2014opennetmon}\cite{tootoonchian2010opentm}) and the controller's real-time workload. Since these two problems are coherent, we approach the optimal results with the joint optimization.

In summary, our paper makes the following contributions:
\begin{itemize}
\vspace{-0.15cm}
  \item We formulate the joint optimization problem as the \emph{Optimal Switch Configuration and Mapping (OSCM)} problem, which aims to keep low communication overhead of controllers by deciding the offline switch control shift based on the switch and controller states in real time.
  \item We provide a rigorous proof of the OSCM problem to be NP-hard and propose a heuristic solution named \emph{\solution} to efficiently solve the problem.
  \item We evaluate the performance of \solution \ under a real topology. Simulation results show that compared with the baseline algorithm, \solution \ can reduce the communication overhead up to 52.6\% during a moderate controller failure that active controllers have enough ability to recover 100\% flows from offline switches, and can reduce the communication overhead up to 61.2\% during a serious controller failure by setting to recover 90\% flows from offline switches.
  \vspace{-0.15cm}
\end{itemize}


The rest of the paper is organized as follows. In Section \ref{sec:motivation}, we introduce the background of SDN and the motivation of this paper. Section \ref{sec:design} introduces our design considerations, and Section \ref{sec:math} mathematically formulates our design as the OSCM problem. Section \ref{sec:solution} proves the OSCM problem's complexity and proposes \solution \ to efficiently solve the problem. We evaluate and analyze the performance of \solution \ in Section \ref{sec:simulation}. Section \ref{sec:relatedworks} introduces related works, and Section \ref{sec:conclusion} concludes this paper.

\begin{figure*}[!ht]
\centering
\subfigure[Network composition]{
\includegraphics[width=2.3in]{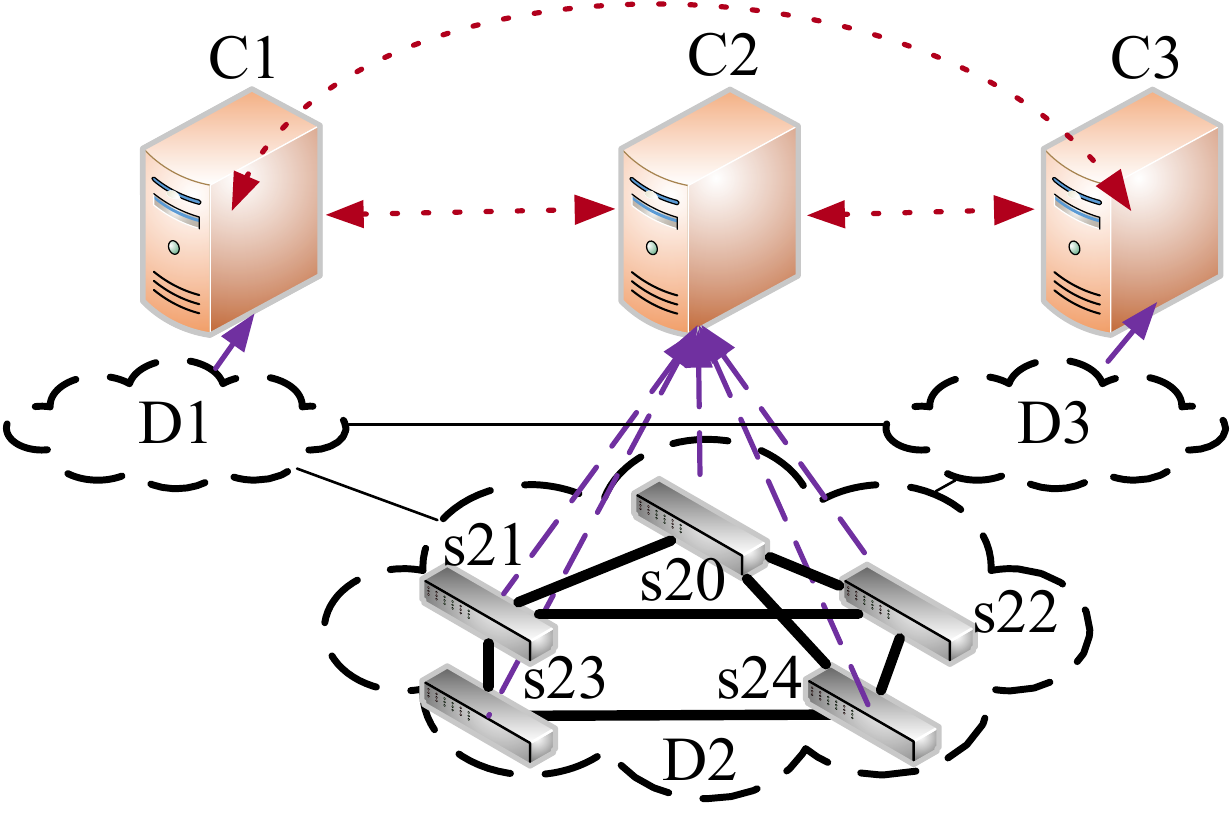}}
\hspace{0.05in}
\subfigure[Flows and paths in domain $D_2$]{
\includegraphics[width=2.1in]{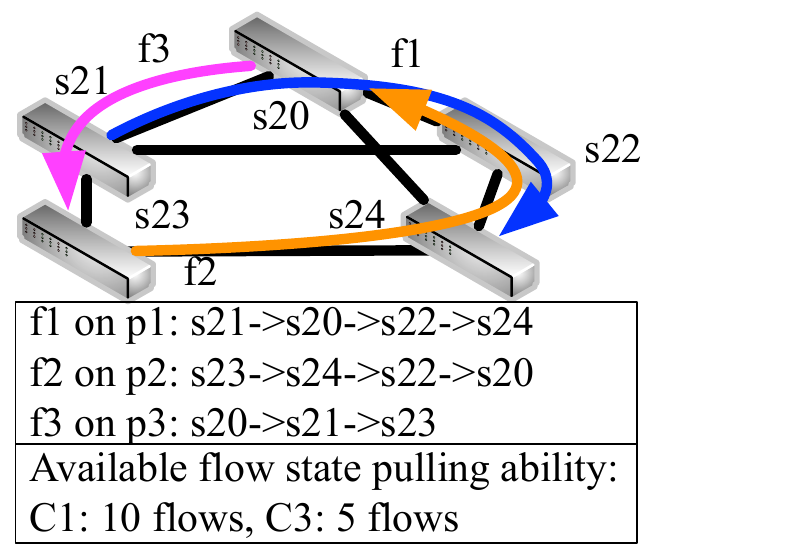}}
\hspace{0.05in}
\subfigure[Controller $C_2$ fails. Switches in $D_2$ should be controlled by other active controllers.]{
\includegraphics[width=2.25in]{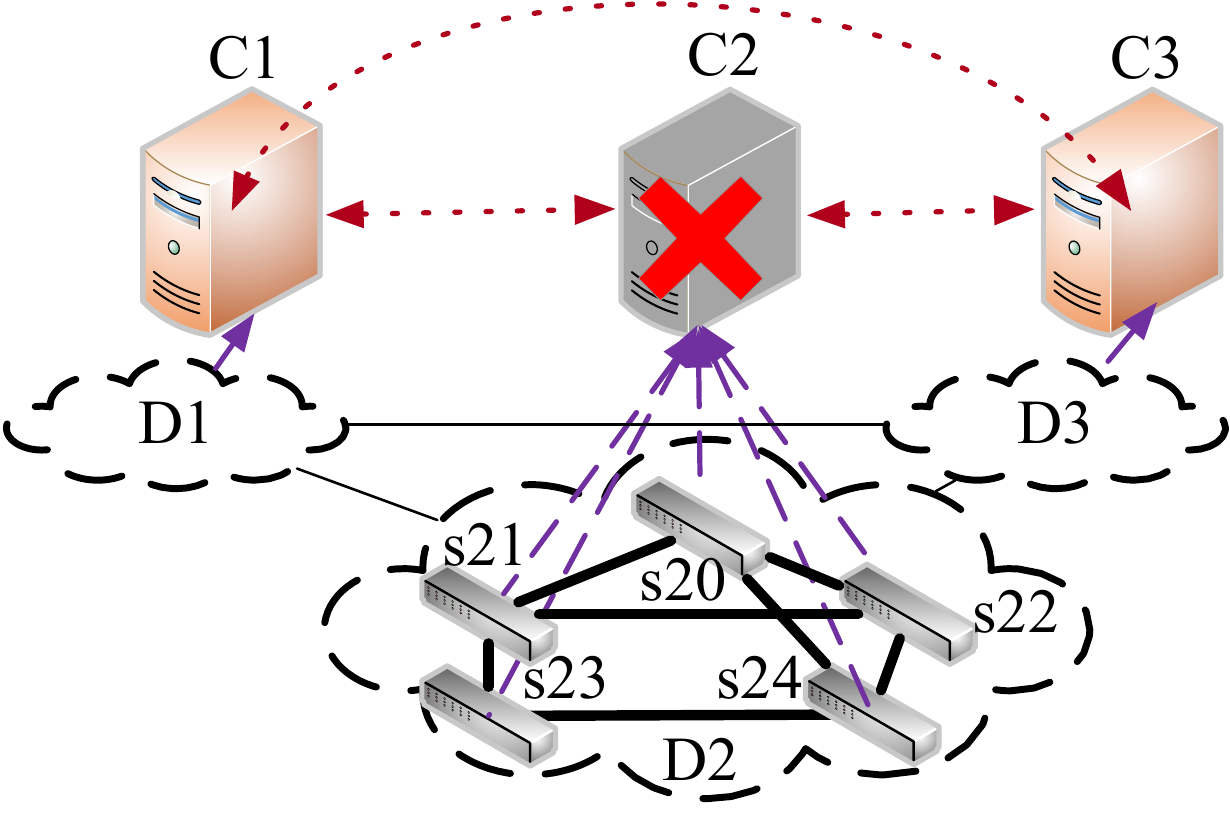}}
\subfigure[Low communication overhead but overwhelming $C_3$. Under $C_2$ failure, switches in $D_2$ are mapped to their nearest active controller $C_3$.]{
\includegraphics[width=2.25in]{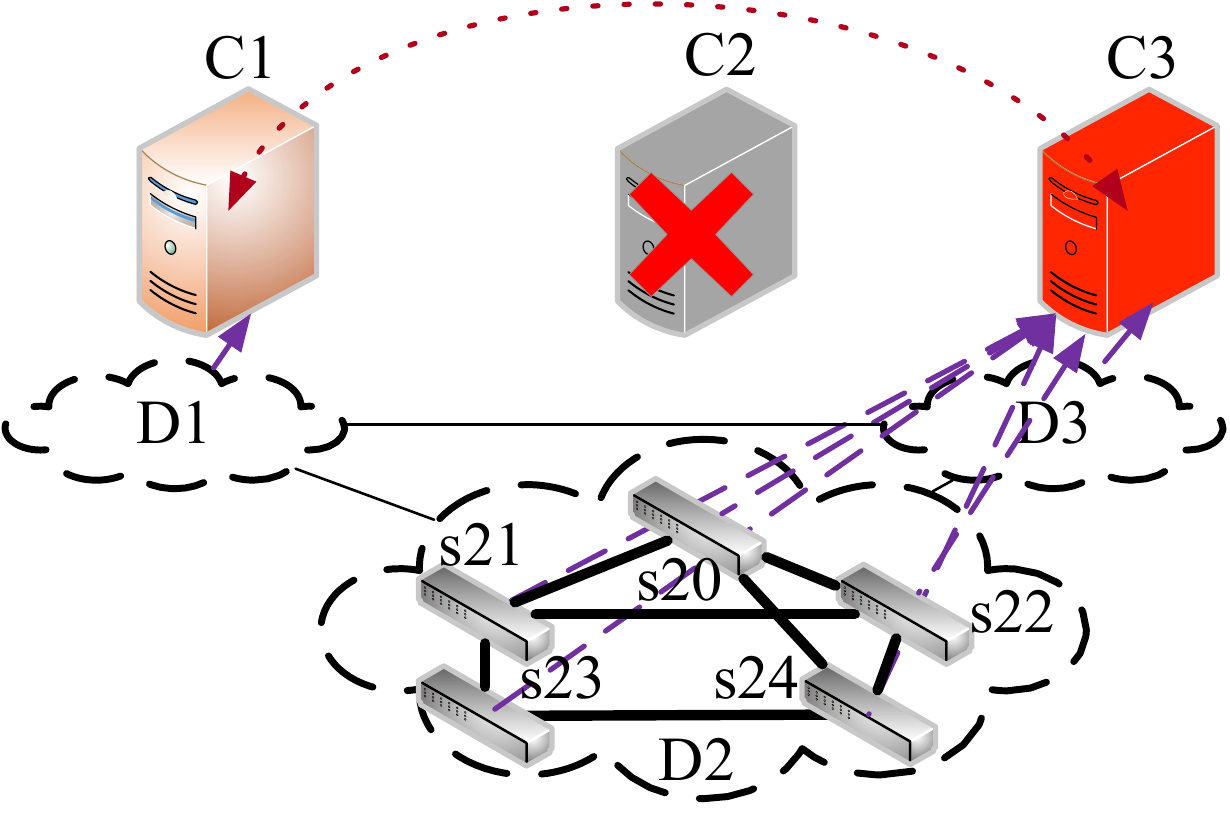}}
\hspace{0.05in}
\subfigure[High communication overhead without overloading active controllers. Under $C_2$ failure, switches in $D_2$ are individually mapped to active controllers by considering the switch-controller propagation delay and the constraint of controllers' processing abilities.]{
\includegraphics[width=2.25in]{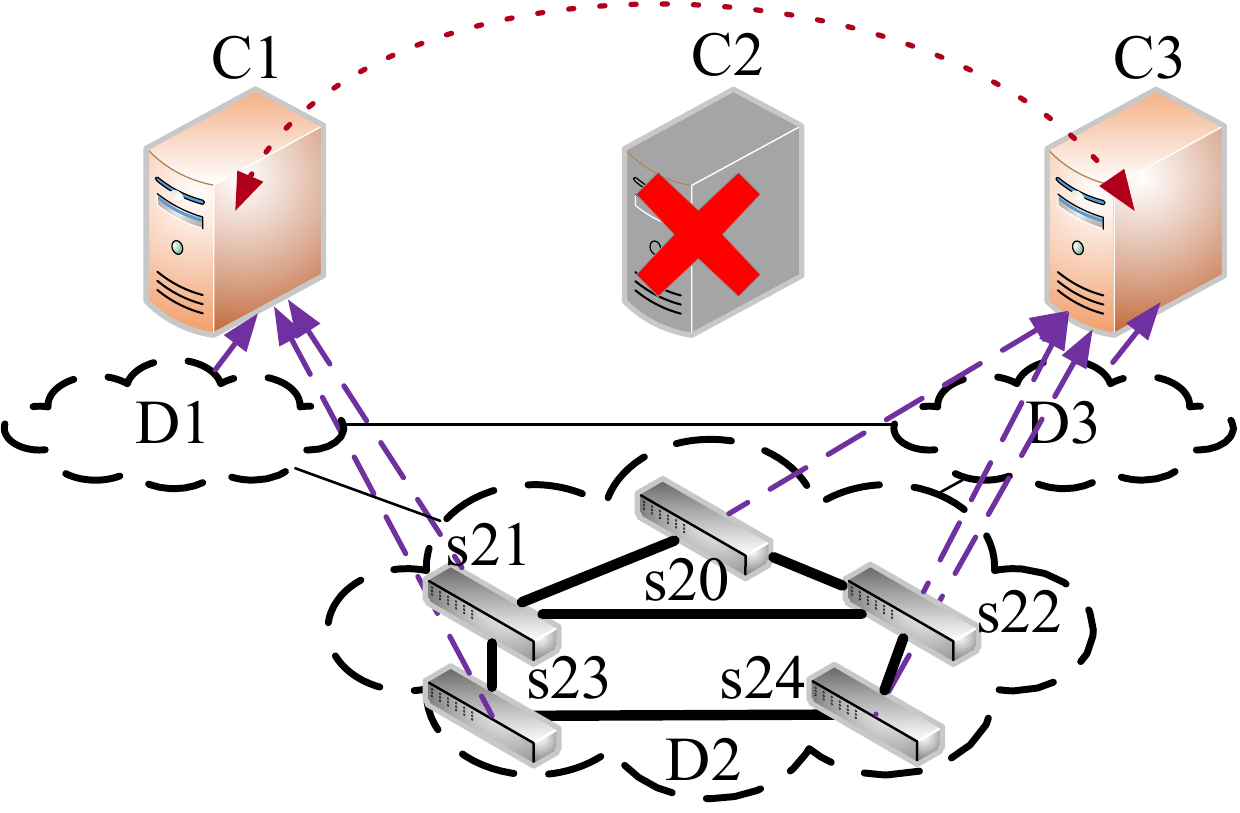}}
\hspace{0.05in}
\subfigure[Legend.]{
\includegraphics[width=2.1in]{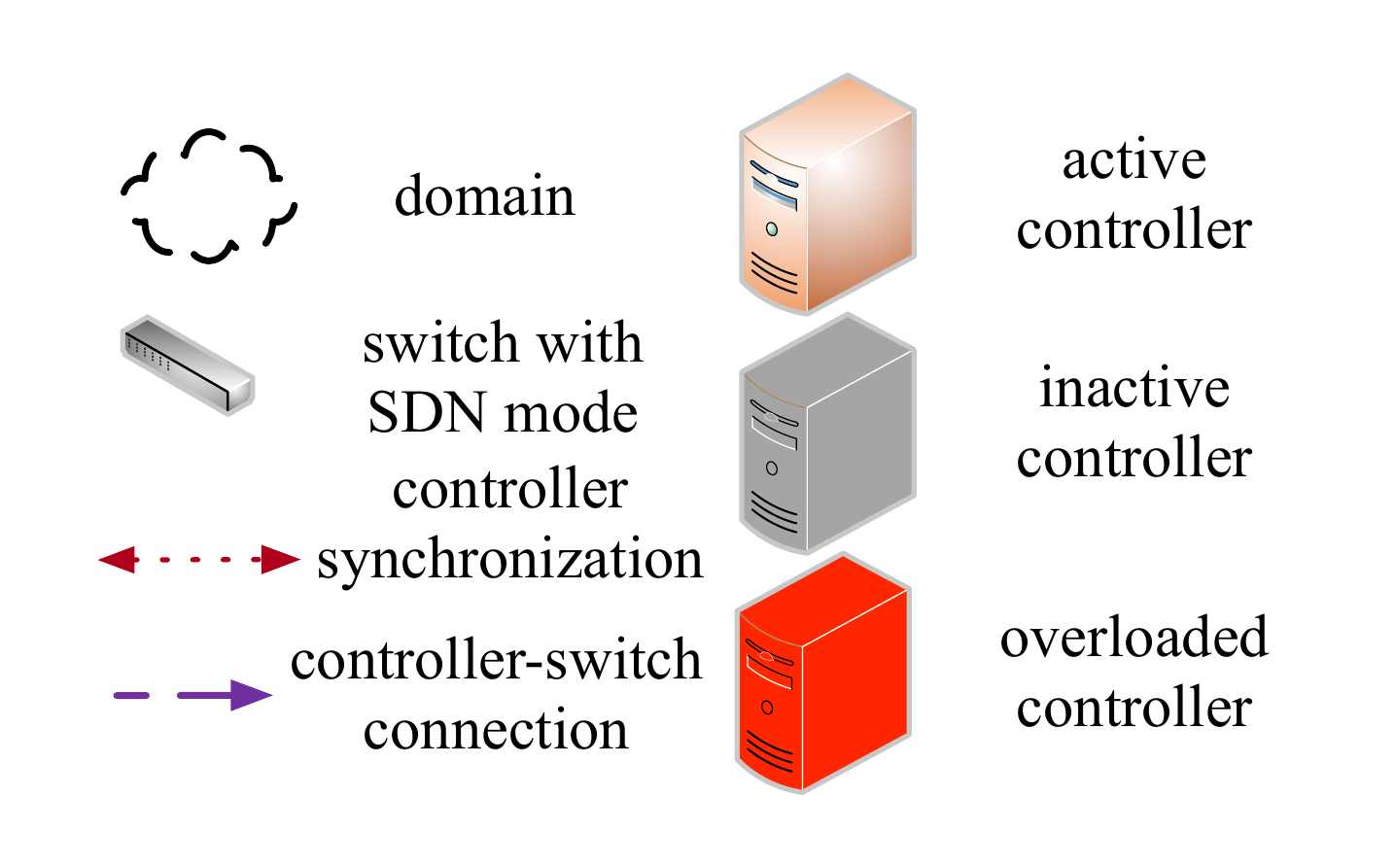}}
\vspace{-0.4cm}
\caption{A motivation example of switch remapping under the controller failure. Switches in $D_2$ are closer to $C_3$ than $C_1$.}
\label{fig:motivation_example}
\vspace{-0.4cm}
\end{figure*}

\section{Background and Motivation}
\label{sec:motivation}
In this section, we introduce the background of SDN, analyze the limitation of SDN under controller failures, and present opportunities to solve the problem using features available in commercial SDN switches. 

\subsection{Blessing of the SDN}
\label{sec:background}
One big benefit of SDN is to provide flexible control on traffic flows based on the global state of the network. To achieve this benefit, the SDN controller can establish forwarding paths for individual flows reactively when they enter the network for the first time or proactively before they arrive at the network. During the network operation, the controller periodically pulls flow state information from the controlled switches to update its global network view and dynamically changes some flows' paths to improve network performance. These unique features and advantages, called \emph{flow programmability}, help SDN to prevail over traditional network techniques. Therefore, many networks start to deploy SDN \cite{uber}\cite{jain2013b4}\cite{att}. 

For an SD-WAN that consists of many switches, we usually divide the WAN into multiple domains with different number of SDN switches and use a logically centralized control on domains with distributed controllers \cite{guo2014improving}. In each domain, its controller can quickly reply to requests from switches and synchronize with other controllers to maintain the consistent network view.


\subsection{Curse of the SDN}
\label{sec:curse} 
The normal operations of an SD-WAN rely on the controllers' decision and the communication between controllers and switches for conducting the decision and pulling flow state information. The controller becomes the Achilles Heel of SDN. In other words, an SDN switch will be out of control if its controller fails. In order to provide a resilient control of the network, an SDN switch usually connects to a master controller and several backup controllers \cite{of}. When the master controller of a switch crashes, its connection to the switch becomes inactive, and the switch will request one of its backup controllers to become its new master controller. We call the switches previously controlled by failed controllers the \emph{offline switches}. The problem of remapping the control of offline switches to other active controllers is called \emph{SDN switch remapping under controller failures}. This switch remapping has two impacts on other active controllers:
\begin{itemize}
    \item{\textbf{Overloading controllers}:}
A master controller mainly has two types of operations on switches: (1) flow entry operations to establish/update flows' forwarding paths and (2) flow state pulling operations to get the network state variation. Both of the two operations consume the processing ability of the master controller. Backup controllers only maintain the connection to their switches without operations until they become master controllers. Becoming the new master controller of some switches from remote domains increases the processing load of a controller, potentially overloading the controller \cite{hu2018adaptive}\cite{yao2013cascading}. Existing studies show the switches' requests handled by an overloaded controller could experience long-tail latency \cite{xie2018cutting}, which could degrade the network performance significantly. 

    \item{\textbf{Increasing the communication overhead of controllers}:}
    The communication overhead of a controller is proportional to two factors: (1) the propagation latency between the controller and its controlled switches and (2) the number of flows in the switches. In WANs, the propagation latency is the dominant factor among all latencies because the propagation latency bounds a controller's control reactions that can be executed at a reasonable speed \cite{heller2012controller}\cite{yao2014capacitated}. A long propagation latency could limit convergence time (e.g., routing convergence, network update). Thus, remapping offline switches to active controllers should consider the propagation latency among the switches and controllers. Otherwise, the reaction of the controllers to dynamic network changes could be delayed.

    
\end{itemize}
\vspace{-0.1cm}

To better illustrate our view, we use a motivation example in Figure \ref{fig:motivation_example} to show that existing solutions under controller failures suffer from the two impacts. In Figure \ref{fig:motivation_example}(a), an SDN consists of three domains, and each domain is controlled by one master controller and connected to two backup controllers. In this example, we mainly focus on domain $D_2$ and do not show details of the other two domains. In domain $D_2$, controller $C_2$ is the master controller that controls five SDN switches $s20$-$s24$. Controllers $C_1$ and $C_3$ are backup controllers of $D_2$. The three controllers synchronize with each other to maintain the consistent network information. Both $C_1$ and $C_3$ know that $D_2$ has five switches and the flow information of $D_2$. We denote the processing ability of a controller as the number of flows. In this example, without interrupting a controller's normal operations, $C_1$ can pull the state of ten flows, but $C_3$ can only pull the state of at most five flows. Figure \ref{fig:motivation_example}(b) shows the flows in $D_2$ and their paths. 

An SDN is vulnerable. In Figure \ref{fig:motivation_example}(c), controller $C_2$ fails, and the control of the five switches in $D_2$ should be transferred to active controllers $C_1$ and $C_3$. As summarized above, two problems are raised in the following cases:



\begin{enumerate}
\item{{Controller overload}:}
Figure \ref{fig:motivation_example}(d) shows that remapping switches in $D_2$ to active controllers only considers the switch-controller propagation delay. In this figure, the five switches are remapped to their nearest, active controller $C_3$. This solution minimizes the communication overhead of the entire network but controller $C_3$ has to pull the states of eleven flows (i.e., $f1$, $f2$, and $f3$ from $s20$, $f1$ and $f3$ from $s21$, $f1$ and $f2$ from $s22$, $f2$ and $f3$ from $s23$, $f1$ and $f2$ from $s24$), which interrupts its normal operations by introducing queueing delays \cite{xie2018cutting}.

\item{{High controller communication overhead}:} Figure \ref{fig:motivation_example}(e) shows that remapping switches in $D_2$ to active controllers considers the switch-controller propagation delay and controllers' processing abilities. In this figure, switches $s20$, $s22$, and $s24$ are remapped to controller $C_3$ while switches $s21$ and $s23$ are remapped to controller $C_1$. This solution prevents controllers from being overloaded but incurs higher communication overhead than the solution in Figure \ref{fig:motivation_example}(d) due to the long propagation delay among offline switches and controllers.
\end{enumerate}

The above two examples show that existing solutions either suffer from the controller overload or high communication overhead, and they cannot solve the problem of SDN switch remapping under controller failures well.

\subsection{Opportunities}
We note that many commercial SDN switches today are hybrid SDN switches (e.g., Brocade MLX-8 PE \cite{PE}) and support two routing modes: legacy mode and SDN mode. In the legacy mode, switches route flows with the destination-based entries in the routing table generated from legacy routing protocols (e.g., OSPF), whereas in the SDN mode, they route flows using OpenFlow table managed by the controller. In other words, a switch with the legacy mode can work without the controller. Thus, we can reduce the controller's processing load by configuring some switches working under the legacy mode.

\begin{figure}[t]
\centering
\includegraphics[width=3in]{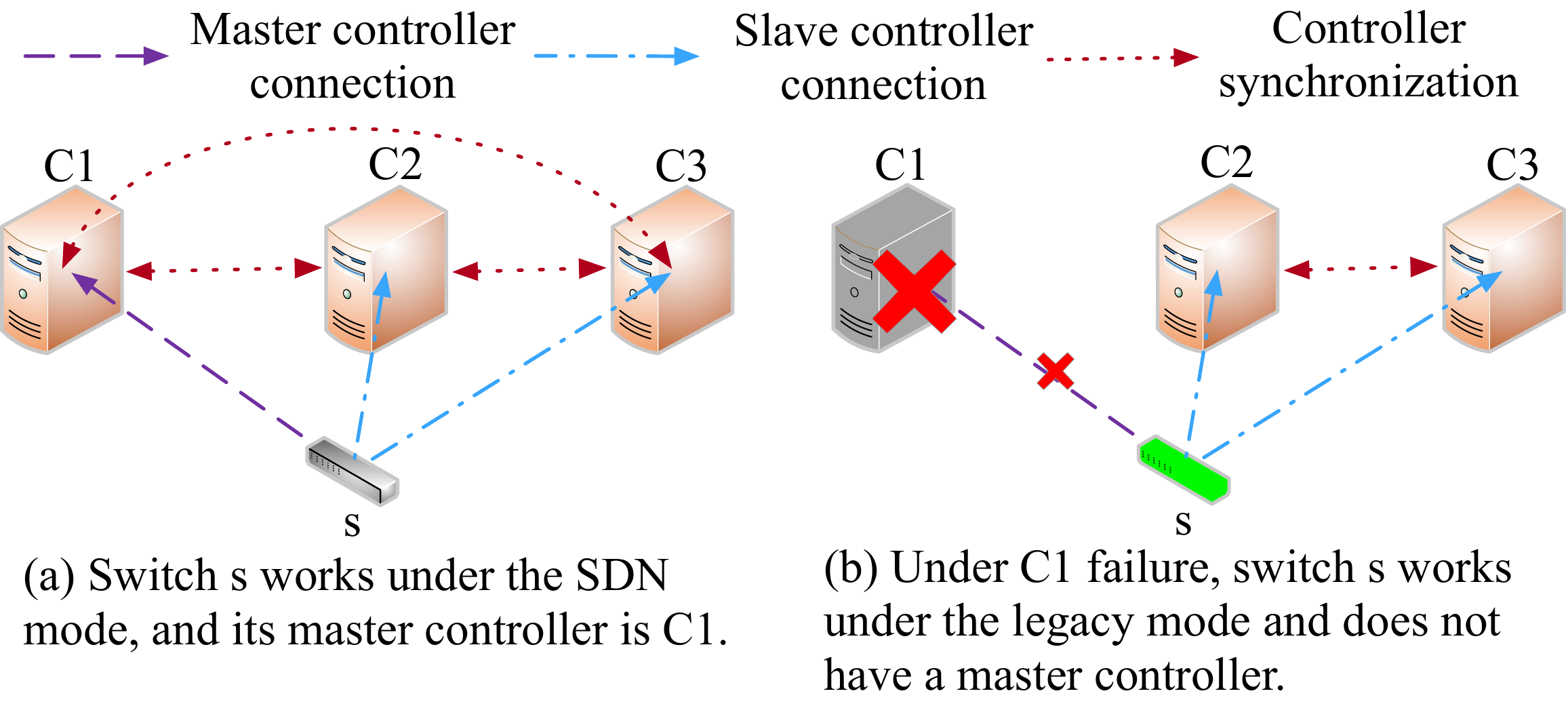}
\vspace{-0.2cm}
\caption{Switch routing mode configuration.}
\vspace{-0.6cm}
\label{fig:comm}
\end{figure}


\begin{figure*}[!ht]
\centering
\subfigure[Realizing the programmable ability of $D_2$. Under $C_2$ failure, $s20$, $s22$, and $s24$ are configured with the legacy mode, and switches $s21$ and $s23$ with the SDN mode are mapped to controller $C_3$.]{
\includegraphics[width=2.3in]{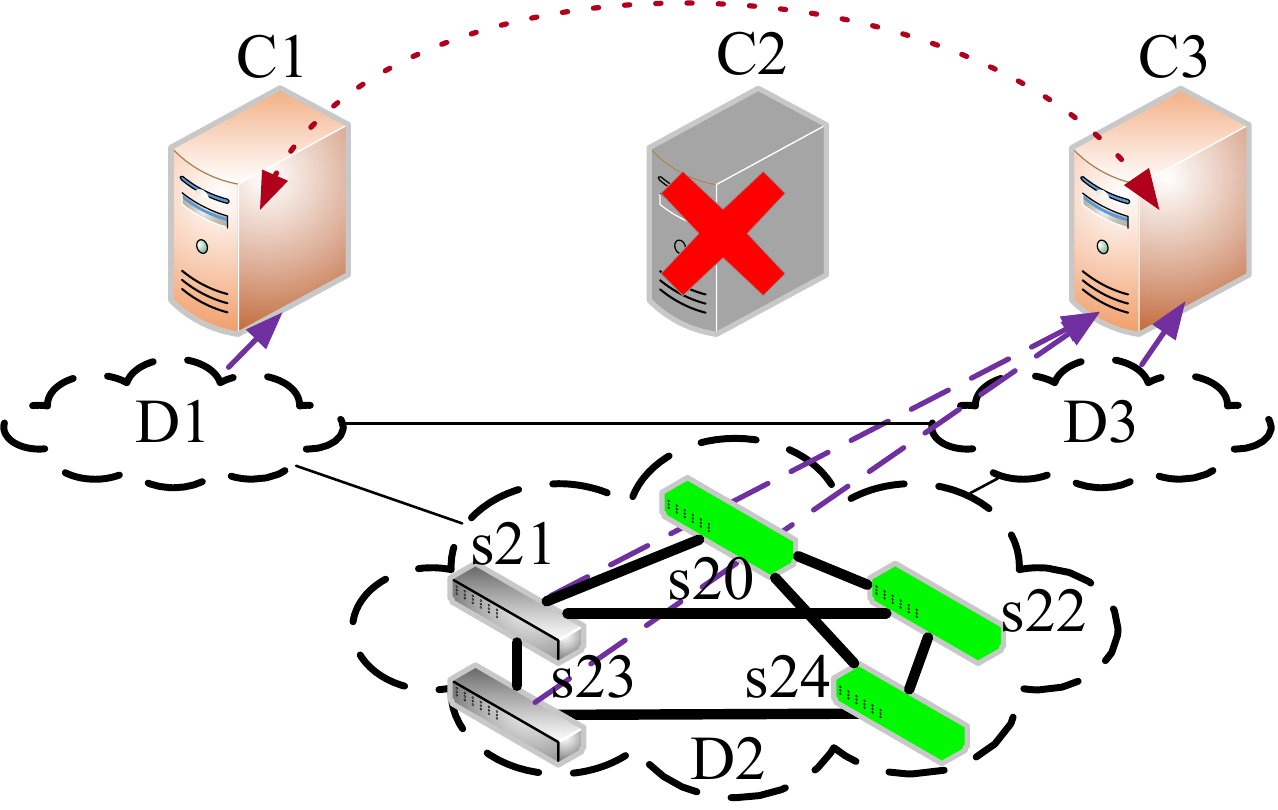}}
\hspace{0.05in}
\subfigure[Joint realizing the programmable ability of $D_2$ and reducing the communication overhead. Under $C_2$ failure, \solution \ configures switches $s21$, $s23$, and $s24$ with the legacy mode and maps switches $s20$ and $s22$ with the SDN mode to their nearest active controller $C_3$.]
{\includegraphics[width=2.3in]{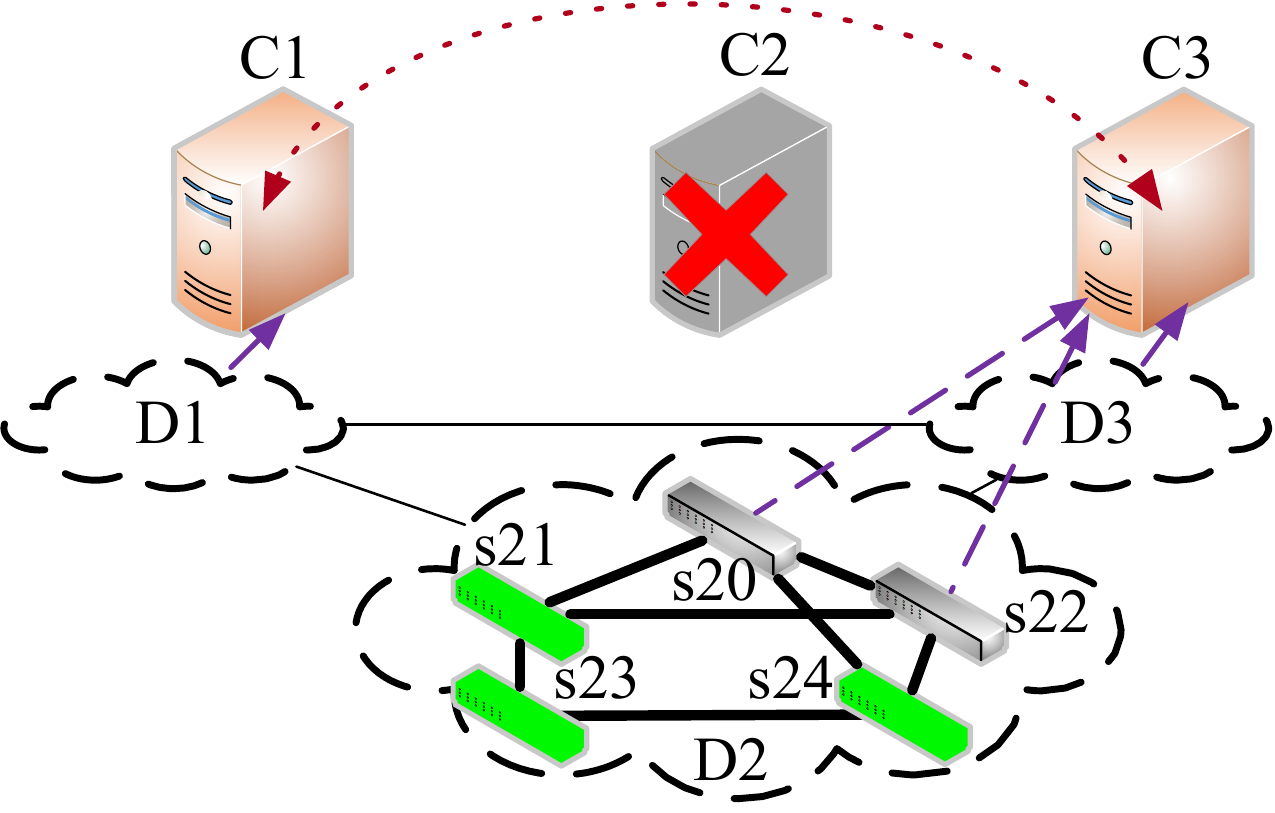}}
\hspace{0.05in}
\subfigure[Legend. Note that the last switch on a flow's forwarding path cannot be used to change a flow's path.]{
\includegraphics[width=2.1in]{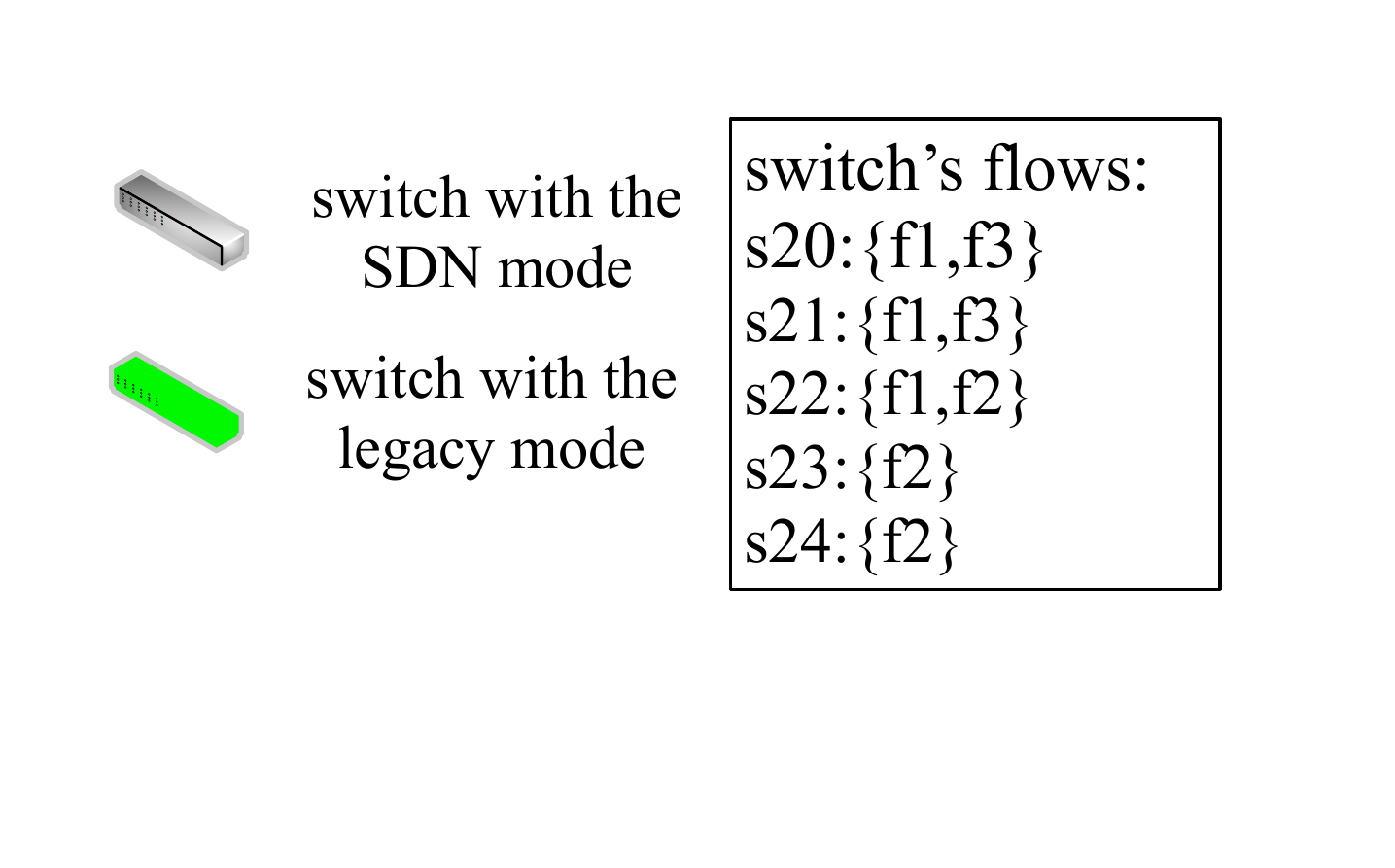}}
\vspace{-0.3cm}
\caption{Switch remapping under a controller failure using the SDN and legacy modes. Switches $s21$ and $s23$ are much far from controllers $C_3$ than switches $s20$, $s22$, and $s24$.}
\label{fig:design}
\vspace{-0.4cm}
\end{figure*}

A controller can dynamically set the routing mode of a hybrid SDN switch by sending a control message. Figure \ref{fig:comm} shows an example of changing the routing mode under a controller failure. In Figure \ref{fig:comm}(a), switch $s$ connects to its master controller $C_1$ and backup controllers $C_2$ and $C_3$. $C_1$ sets $s$ to work under the SDN mode, and the three controllers synchronize with others. In Figure \ref{fig:comm}(b), when $C_1$ fails, $s$ identifies its connection to $C_1$ become inactive and then sends a master controller selection request to $C_1$, $C_2$, and $C_3$. Both $C_2$ and $C_3$ reply a rejection message to switch $s$'s request with a legacy mode configuration, and then $s$ starts to work under the legacy mode. $s$ can go back to the SDN mode when one controller wants to become its new master controller. This dynamic routing mode configuration feature offers us more flexibility to tackle the problem of switch remapping under controller failures.


\section{Design Considerations} 
\label{sec:design}
In this section, we propose to relieve the controller's control cost of offline switches and reduce the communication overhead during controller failures by introducing the optimal switch configuration and mapping problem. 



\subsection{Switch mode configuration problem}
Inspired by the hybrid SDN switches available in the market, we propose to relieve the controller's control cost of offline switches by \emph{degrading} a pure SDN of the offline switches to a hybrid SDN that consists of switches with the SDN mode and legacy mode. In other words, we configure a subset of offline switches to run under the SDN mode and others to run under the legacy mode without the management of controllers. However, when configuring the routing mode on switches, we should guarantee the programmability of flows in the hybrid SDN. 

The programmability of flows is an essential feature of SDN \cite{agarwal2013traffic}\cite{poularakis2017one}\cite{8672633}. Taking the routing problem as an example, the programmability of a flow is the ability to change the flow's path. Existing pure SDN designs realize the network programmability with \emph{a per-hop programmable routing}, which enables the controller to change each flow's path from any switches\footnote{If a switch cannot reach a flow's destination through two paths, we will not change the flow's path from this switch.} on the flow's path. In contrast, by maintaining the basic programmability, we keep only \emph{1-hop programmable routing} for flows of offline switches. In other words, we can still change the path of a flow from offline switches by controlling one switch on the flow's path while letting other switches on the flow's path just forward the flow using the legacy destination-based routing. Thus, the switches using the legacy routing do not need to consult the controllers for routing flows.

Figure \ref{fig:design}(a) shows an example that under the same failure case of Figure \ref{fig:motivation_example}(c), we achieve the programmablity of flows $f1$-$f3$ using switches with the SDN mode and legacy mode. In this figure, switches $s21$ and $s23$ work under the SDN mode, and switches $s20$, $s22$, and $s24$ work under the legacy mode. Flows $f1$, $f2$, and $f3$'s forwarding paths can be changed at switches $s21$, $s23$, and $s21$, respectively.


\vspace{-.3cm}
\subsection{Switch remapping problem} 
With the reduced number of switches under the SDN mode, we have more freedom to remap these switches to active controllers. The switch remapping should maintain low communication overhead between these switches and active controllers without exceeding the controller's processing ability. The load of a controller is defined as the number of flows the controller can control, and it is restricted to the processing ability of the controller. As explained in Section \ref{sec:curse}, the load consists of two parts: flow entry operations and flow state pulling operations. The former one has been considered in many existing works \cite{wang2016dynamic}, while we argue the latter one is more important in the WAN scenario. In WANs, each flow's path is usually proactively configured since each flow is an aggregated large flow of multiple flows and always has a traffic rate. Only a limited number of flow entry operations are conducted to reroute some flows at some extreme situations (e.g., congestion). However, each controller conducts the flow state pulling operations periodically (e.g., every few seconds \cite{xu2017minimizing}) to maintain the updated network view. Per-flow pulling \cite{van2014opennetmon}\cite{tootoonchian2010opentm} are popular state pulling methods. The controller sends a request to a switch to pull the state of one flow. The number of flow state pulling requests is proportional to the number of flows in a switch. Thus, in WANs, the flow state pulling is a big overhead for the controllers and is our main concern in this paper. 

We name the solution that sequentially solves the above two problems one by one the two-phase solution, which obtains the final result by solving the switch mapping problem using the result of the switch mode configuration problem. Figure \ref{fig:design}(a) shows the result of the two-phase solution under the same failure case of Figure \ref{fig:motivation_example}(c). In this figure, switches $s21$ and $s23$ work under the SDN mode and are mapped to controller $C_3$. $C_3$ pulls the state of flows $f1$, $f2$, and $f3$ from switches $s21$, $s23$, and $s21$, respectively.


\vspace{-.9cm}
\subsection{A joint optimization problem}
\label{sc:explanation}
However, it is not enough to consider the above two problems independently because the two problems are correlated. Recall the communication overhead of a controller equals to the propagation latency between the controller and its controlled switches multiplied with the number of flows in the switches. Thus, to achieve the low communication overhead, it may be better to choose more switches (or switches with more flows) with lower delays than to choose fewer switches (or switches with fewer flows) with higher delays.  

Figure \ref{fig:design}(b) shows the result of \solution \ under the same failure case of Figure \ref{fig:motivation_example}(c). In this figure, \solution \ remaps switches $s20$ and $s22$ to controller $C_3$ and configure switches $s21$, $s23$, and $s24$ under the legacy mode. Flows $f1$, $f2$, and $f3$'s forwarding paths can be changed at switches $s20$ and $s22$, at switch $s22$, and at switch $s20$, respectively. Under the given processing ability of controllers, \solution \ outperforms the two-phase solution in Figure \ref{fig:design}(a) in two aspects. First, \solution \ has a higher flow programmablitily since flow $f1$ traverses two switches with the SDN mode. Second, \solution \ has a lower communication overhead. \solution \ pulls flow $f1$ twice from two switches, but switches $s20$ and $s22$ are much closer to controller $C3$ than switches $s21$, $s23$, and $s24$. Thus, for the communication overhead, the significant decrease of propagation delay compromises the increased number of flows. 

Therefore, we should jointly consider the number of flows in offline switches and the propagation delay among the switches and active controllers to configure switches' routing mode and the controller-switch mapping. We name this problem \emph{Optimal Switch Configuration and Mapping (OSCM)} problem and literally explain it as follows: \emph{under controller failures, we need to maintain the programmability of flows from offline switches with the minimum communication overhead among offline switches and active controllers by efficiently configuring each offline switch with a routing mode and mapping the offline switches with the SDN mode to the active controllers}.  


\section{Problem Formulation}
\label{sec:math}
In this section, we introduce how to optimally configure and remap switches by modeling the system, introducing constraints and the objective function, and finally formulating an optimization problem. For simplicity, in the rest of this paper, we use a switch instead of an offline switch.

\subsection{System description}
Typically, an SD-WAN consists of $H$ controllers at $H$ locations, and each controller controls a domain of switches. Controllers $C_{M+1}$, ... , $C_H$ fail, and they control $N$ switches in total. The set of active controllers is $\mathcal{C} = \{C_{1}, ..., C_j, .... C_M\}$, and the set of switches controlled by the failed controllers are $\mathcal{S} = \{s_1, ..., s_i, ..., s_{N}\}$. We need to select some switches from $\mathcal{S}$, configure them with the SDN mode, and map these selected switches to controllers in $\mathcal{C}$; the rest of switches in $\mathcal{S}$ are configured with the legacy mode. If switch $s_i \in \mathcal{S}$ is configured with the SDN mode, $x_{i} = 1$; otherwise, the switch is configured with the legacy mode and does not rely any controller, and thus $x_{i} = 0$. We use $z_{ij} = 1$ to denote that switch $s_i \in \mathcal{S}$ under the SDN mode is mapped to controller $C_j \in \mathcal{C}$; otherwise $z_{ij} = 0$. Since both $x_i$ and $z_{ij}$ are binary variables, and a feasible switch-controller mapping requires that a switch is under the SDN mode and mapped to an active controller, we can have 
\begin{equation}
\label{eq:switch}
x_i * z_{ij} = z_{ij}, \forall i\in[1,N],\ \forall j\in[1,M].
\end{equation}
The set of flows from switches $\mathcal{S}$ is $F = \{f^1, f^2, ..., f^l, ..., f^{L}\}$. If flow $f^l$'s forwarding path traverses switch $s_i$, and $s_i$ has at least two paths to $f^l$'s destination, we have $\beta_{i}^l = 1$, otherwise $\beta_{i}^l = 0$.  If flow $f^l$ is a programmable flow, we have $y^l$ = 1, otherwise $y^{l} = 0$. 



\subsection{Constraints}
\subsubsection{Switch-controller mapping constraint} 
If switch $s_i$ is configured with the SDN mode, it must be controlled by only one controller; if switch $s_i$ is configured with the legacy mode, it is not controlled by any controller. Thus, we have  
 \begin{equation}
\sum_{j=1}^{M} z_{ij} = x_{i}, \forall i\in[1,N].
\label{eq:mapping}
\end{equation}

\subsubsection{Controller processing ability constraint} 
If some controllers fail, it is unfair to overload other active controllers to take full responsibility for the failed controllers to control offline switches. Active controllers should only try their best to control the offline switches without interrupting their normal operations. The state pulling operations of a switch equal to the total number of flows in the switch's flow table. We measure a controller's processing ability as the number of flows that the controller can normally pull from its controlled switches without introducing extra delays (e.g., queueing delay). The processing load of a controller should not exceed the controller's processing ability. It can be written as
\begin{equation*}
\sum_{i=1}^{N}(\sum_{l = 1}^{L} \beta^l_{i}* x_i* z_{ij})  \leq A_j^{rest}, \forall j\in[1,M],
\end{equation*}
where $A_j^{rest}$ denotes the available processing ability of controller $C_j$.
Bringing \eqref{eq:switch} into the above inequality and letting $g_i $ denote the number of flows in switch $s_i$
\begin{equation}
\label{eq:total_flow}
g_i = \sum_{l = 1}^{L} \beta^l_{i}, ~ \forall i\in [1,N],
\end{equation}
we can reformulate the above nonlinear constraints as the following linear constraints: 
\begin{equation}
\sum_{i=1}^{N}(g_i  * z_{ij}) \leq A_j^{res}, \forall j\in[1,M].
\label{eq:ability}
\end{equation}

\subsubsection{Flow programmability constraint} 
If a switch works under the SDN mode, the flows in the switch become programmable. The flow $f^l$'s programmability can be expressed as follows: 
\begin{equation}
y^l \leq  \sum_{i = 1}^{N} (\beta^l_{i} *x_{i}), \forall l\in [1,L].
\label{eq:program_flow}
\end{equation}
In the above inequality, the equal sign comes when there is only one offline switch with the SDN mode that contains flow $f^l$. If multiple switches contain this flow, the inequality sign is used.
%

The flow programmability equals the total number of unique programmable flows. If we require $Q$ unique flows are programmable, we have 
\begin{equation}
  Q \leq \sum_{l = 1}^{L} y^l.
  \label{eq:program_network}
\end{equation}

%
%
%

\subsection{Objective function}


The objective is to minimize the communication overhead of active controllers to pull flow state from offline switches, which equals the total propagation delay of programmable flows between the switches with the SDN mode and their newly mapped controllers. We use $D_{ij}~(D_{ij} \geq 0)$ to denote the propagation delay between switch $s_i$ and controller $C_j$ and formulate the overhead as follows:
%
\begin{equation*}
obj= \sum_{j = 1}^{M}\sum_{i = 1}^{N} (g_i* D_{ij}* z_{ij}).
\end{equation*}
If we use $w_{ij}$ to denote controller $C_j$'s communication overhead to switch $s_i$:
\begin{equation}
w_{ij} = g_i* D_{ij}, \forall i\in[1,N], \forall j\in[1,M],
\end{equation}
we can write the objective function as follows
\begin{equation}
 obj= \sum_{j = 1}^{M}\sum_{i = 1}^{N} (w_{ij}  *z_{ij}).
\end{equation}

\subsection{Problem formulation}

The goal of our problem is to minimize the communication overhead between active controllers in $\mathcal{C}$ and offline switches in $\mathcal{S}$ and provide the programmability for flows in $F$ by smartly configuring switches in $\mathcal{S}$ and mapping switches with the SDN mode to active controllers in $\mathcal{C}$. Therefore, we formulate the OSCM problem as follows:
\begin{equation}
\tag{P}
\label{problem}
\begin{aligned}
& \underset{z,y}{\text{min}}
& & \sum_{j = 1}^{M}\sum_{i = 1}^{N} (w_{ij} * z_{ij})& \\
& \text{s.t.} & & \eqref{eq:mapping}\eqref{eq:ability}\eqref{eq:program_flow}\eqref{eq:program_network},  &\\
& & & z_{ij}, y^l \in \{0,1\}, \\
& & & \forall i\in[1,N],\forall j\in[1,M],\forall l\in [1,L],   
\end{aligned}
\end{equation}
where $\{w_{ij}\}$, $\{g_i\}$, $\{\beta^l_{i}\}$, and $\{A_j^{rest}\}$ are constants, and $\{z_{ij}\}$ and $\{y^l\}$ are design variables. In the OSCM problem, the objective function is linear, and variables are binary integers. Thus, this problem is an integer programming.  



\section{Solution}
\label{sec:solution}
In this section, we first analyze the complexity of the OSCM problem and then propose our \solution \ algorithm for solving the problem. 

\subsection{Complexity analysis}
\begin{mythe}
	\label{the:nphard}
	For a special case with two conditions: (1) all flows from offline switches should be programmable, and (2) each flow traverses only two switches and has different source and destination switches with others, the OSCM problem is NP-hard.
\end{mythe}
\noindent\textbf{Proof: }	We first introduce the \emph{Generalized Assignment Problem (GAP)} \cite{gary1979computers}. The GAP aims to minimize the cost assignment of $n$ tasks to $m$ agents such that each task is precisely assigned to one agent subject to capacity restrictions on the agents. A typical formulation of the GAP is shown below:
\begin{equation}
\label{problem:gap}
\begin{aligned}
& \underset{x}{\text{min}}
& & \sum_{j = 1}^{m}\sum_{i = 1}^{n} (c_{ij} * x_{ij})& \\
& \text{s.t.} & &
\sum_{i=1}^{n}(a_{ij}  * x_{ij}) \leq b_j, \forall j\in[1,m], &\\
& & & \sum_{j =1}^{m}x_{ij} = 1, \forall i\in [1,n], &\\
& & & x_{ij} \in \{0,1\},\forall i\in[1,n],\forall j\in[1,m],               
\end{aligned}
\end{equation}
where $c_{ij}$ is the cost of assigning task $i$ to agent $j$, $a_{ij}$ is the capacity of task $i$ when the task is assigned to agent $j$, and $b_j$ is the available capacity of agent $j$. Binary variable $x_{ij}$ equals 1 if task $i$ is assigned to agent $j$, otherwise it equals 0. It has been proved when assigning multiple tasks to an agent and ensuring each task is performed exactly by one agent, the GAP is NP-hard \cite{gary1979computers}.




We then prove for a special case of conditions (1) and (2), problem \eqref{problem} and the GAP are equivalent problems. Given condition (1) that all flows from offline switches should be programmable, we have $Q = L$. \eqref{eq:program_network} can be changed to $y^l = 1$ for all $l\in[1,L]$, and \eqref{eq:program_flow} can be rewritten as follows:
\begin{equation}
1 \leq  \sum_{i = 1}^{N} (\beta^l_{i} * x_{i}), \forall l\in [1,L].
\end{equation}
Recall a flow cannot change its path at its destination switch. Given condition (2) that each flow traverses only two switches and has different source and destination switches with others, we have that each offline switch has a unique flow, and the number of offline switches equals the number of unique flows. That is $\beta_{i_0}^{i_0} = 1$ for a specific $i_0\in [1,N]$. Thus, we can change the above inequality as the following equation 
\begin{equation*}
1 = \sum_{i = 1}^{N} (\beta^l_{i} * x_{i}) =  \beta^{i_0}_{i_0} * x_{i_0} = x_{i_0}, \forall i_0\in [1,N].
\end{equation*}

Bringing the above equation into \eqref{eq:mapping}, we have 
 \begin{equation}
\sum_{j=1}^{M} z_{ij} =1, \forall i\in[1,N].
\label{eq:mapping_new}
\end{equation}

Following the above two conditions, our OSCM problem can be reformulated as follows:
\begin{equation}
\tag{P'}
\label{problem:simple}
\begin{aligned}
& \underset{z}{\text{min}}
& & \sum_{j = 1}^{M}\sum_{i = 1}^{N} (w_{ij} * z_{ij})& \\
& \text{s.t.} & & \eqref{eq:ability} \eqref{eq:mapping_new},  &\\
& & & z_{ij} \in \{0,1\},\forall i\in[1,N],\forall j\in[1,M].               
\end{aligned}
\end{equation} 

\begin{table}[!t]
\centering
\caption{Notations}
\vspace{-.3cm}
\label{table:notation}
\begin{tabular}{|p{1cm}|p{7.1cm}|}
\hline
Notation     & Meaning                                                                                                                                        \\ \hline
$\mathcal{S}$				& 		the set of offline switches, $\mathcal{S} = \{ s_i \ |\ i \in [1, N] \}$				\\ \hline
$\mathcal{W}(i)$ & the communication overhead of switch $s_i$, $\mathcal{W}(i) = \{w_{i1},...,w_{ij},...,w_{iM}\}, i \in [1,N]$                                     \\ \hline
$\mathcal{C}(i)$				& 		the set of active controllers by sorting $\mathcal{C} = \{ C_j \ |\ j \in [1, M] \}$ following the ascending order of $\mathcal{W}(i)$,  $i \in [1,N]$		\\ \hline
$\mathcal{A}$           & the set of the available processing capacity of controllers, $\mathcal{A} = \{A^{rest}_j\ |\ j \in [1, M] \}$           \\ \hline

%
$\mathcal{G}$ & the number of flows in switches, $\mathcal{G} = \{g_i\ |\ i\in[1, N] \}$                                   \\ \hline
%
%

$\mathcal{B}$ & the set of flow-switch relationship, $\mathcal{B} = \{B_{1},...,B_i,...B_{N}\ |\ i \in [1,N]\}$, $B_i =\{\beta_{i}^1,...,\beta_{i}^l,...,\beta_{i}^L\}$\\ \hline
$\mathcal{X}$ & the set of offline switches with the SDN mode, $\mathcal{X} = \{i\in[1, N] \ |\ x_i =1\}$                                   \\ \hline
$\mathcal{Z}$ & the set of the mapping relationship between offline switches with the SDN mode and active controllers, $\mathcal{Z} = \{(i,j)\in [1, N] \times [1,M] \ |\ z_{ij} =1\}$ \\ \hline
$\mathcal{Y}$ & the set of controllable flows, $\mathcal{Y} = \{l\in[1, L] \ |\ y_l =1\}$                                   \\ \hline

$\delta$	& 	a number that indicates the maximum number of flows that are different from existing programmable flows		\\ \hline

\end{tabular}
\vspace{-0.4cm}
\end{table}

Problem \eqref{problem:simple} aims to minimize the communication cost of $N$ switches to $M$ controllers such that each switch is precisely assigned to one controller subject to processing ability restrictions on the controllers. We can treat switch $s_i$ and controller $C_j$ in problem \eqref{problem:simple} as task $i$ and agent $j$ in the GAP. By this construction, it is easy to prove that there exists the minimum communication cost by mapping switches in $\mathcal{S}$ to controllers $\mathcal{C}$, if and only if there exists the optimal solution of the GAP by assigning $n$ tasks to $m$ agents. The construction can be done in polynomial time. In problem \eqref{problem:simple}, the mapping between switches and controllers could be many to one. Since the GAP is NP-hard when multiple tasks are assigned to an agent, and each task is performed exactly by one agent \cite{gary1979computers}, problem \eqref{problem:simple} is NP-hard. \qedd

Problem \eqref{problem:simple} is a special case of the OSCM problem and is NP-hard. Therefore, we can have the following conclusion:
\vspace{-.15cm}
\begin{mythe}
	\label{the:nphard}
	The OSCM problem is NP-hard.
\end{mythe}


\begin{algorithm}[!t]
\caption{\solution}
\label{alg}
\begin{flushleft}
	{\bf{Input:}} $\mathcal{S}$, $\mathcal{C}(i)$, $\mathcal{A}$, $\mathcal{G}$, $\mathcal{B}$;\\
	{\bf{Output:}} $\mathcal{X}$, $\mathcal{Z}$, $\mathcal{Y}$;
\end{flushleft}
    
\begin{algorithmic}[1]
	\State $\mathcal{X}=\emptyset$, $\mathcal{Z}=\emptyset$, $\mathcal{Y}=\emptyset$;
		
	\While {True}
		\State $\delta$ = 0, $i_0$ = NULL, $j_0 $ = NULL;		
		\State //find the switch with the maximum number of flows that are different from existing programmable flows;
		\For {$s_{i} \in \mathcal{S}$}
			\For {$l \in\{\beta_i^l = 1, l \in[1,L]\}$ }
				\If {$ l \in \mathcal{Y}$}
					\State $\beta_i^l = 0$;
				\EndIf
			\EndFor
			\If {$|\sum_{l=1}^L \beta_i^l| > \delta$ }
				\State $\delta = |\sum_{l=1}^L \beta_i^l|$, $i_0 = i$;    
			\EndIf
		\EndFor
		\State //assign switch $s_{i_0}$ to controller $C_{j_0}$, which has the lowest communication overhead and enough processing ability 
		\For{$C_{j} \in\mathcal{C}(i_0)$}
		\If {$A_{j}^{rest} - g_{i_0} \geq 0$}
				\State $j_0 = j$, $\mathcal{X} \leftarrow \mathcal{X}\cup {i_0}$, $\mathcal{{Z}} \leftarrow \mathcal{{Z}}\cup {(i_0,j_0)}$;
			\State $A_{j_0}^{rest}= A_{j_0}^{rest} - g_{i_0}$;
				\For {$l \in\{\beta_i^l = 1, l \in[1,L]\}$}
						\State	$\mathcal{Y} \leftarrow \mathcal{Y}\cup l$;
				\EndFor
		\State break;
		\EndIf
		\EndFor
		\State $\mathcal{S} \leftarrow \mathcal{S} \setminus s_{i_0}$;
		\If  {$|\mathcal{S}| == \emptyset$ or $|\mathcal{Y}| \geq Q$}
			\State break;
		\EndIf
	\EndWhile
\State return $\mathcal{X}, \mathcal{Z}, \mathcal{Y}$;
\end{algorithmic}
\end{algorithm}
\vspace{-0.4cm}

\subsection{\solution \ algorithm}
\label{\solution}
Typically, we can use existing integer program optimization solvers to obtain the OSCM problem's optimal solution. However, for the problem with a large network, the solver could require a very long time or sometimes is impossible to find a feasible solution. Therefore, we propose a heuristic algorithm called \solution \ for solving the problem to achieve the trade-off between the performance and time complexity.




The idea behind \solution \ is to select and test variables based on their importance. The first priority of our problem is to enables many unique flows from offline switches to become programmable flows. Thus, we first select a switch that has the maximum number of flows which are different from existing programmable flows. This switch selection method helps us to efficiently rescue as many unique flows as possible in each iteration. For this selected switch, we choose a switch-controller mapping among all mappings in the ascending order of the communication overhead and then test whether the mapping satisfies the controller's processing ability. If yes, the mapping is selected, and all flows in the switch become programmable; otherwise, a new mapping is tested. This mapping selection method effectively reduces the communication overhead.


\begin{table*}[ht]
	\caption{Default relationship between controllers, switches, and the number of flows in the switches under \topoa \ topology.}
	\label{placement}
	\centering	
\begin{tabular}{|p{2cm}|p{0.3cm}|p{0.2cm}|p{0.3cm}|p{0.2cm}|p{0.2cm}|p{0.3cm}|p{0.2cm}|p{0.2cm}|p{0.2cm}|p{0.2cm}|p{0.2cm}|p{0.2cm}|p{0.2cm}|p{0.2cm}|p{0.2cm}|p{0.3cm}|p{0.2cm}|p{0.2cm}|p{0.2cm}|p{0.3cm}|p{0.2cm}|p{0.2cm}|p{0.3cm}|p{0.2cm}|p{0.2cm}|}
\hline

Controller ID & \multicolumn{4}{c|}{2} & \multicolumn{4}{c|}{5} & \multicolumn{4}{c|}{6} & \multicolumn{5}{c|}{13} & \multicolumn{2}{c|}{20} & \multicolumn{6}{c|}{22}       \\ \hline
Switch ID          & 2    & 3   & 9    & 16 & 4   & 5    & 8   & 14  & 0    & 1   & 6   & 7   & 10 & 11 & 12 & 13  & 15 & 19         & 20         & 17  & 18 & 21 & 22  & 23 & 24 \\ \hline
Number of flows       & 127  & 71  & 121  & 57 & 49  & 153  & 53  & 61  & 81   & 49  & 77  & 93  & 65 & 59 & 71 & 225 & 67 & 49         & 61         & 133 & 49 & 67 & 111 & 49 & 57 \\ \hline

\end{tabular}
\vspace{-0.2cm}
\end{table*}

\newcommand{\length}{1.6in}
\begin{figure*}[t]
\centering
\subfigure[Percentage of programmable flows from offline switches. The higher, the better.]{
\includegraphics[width=\length]{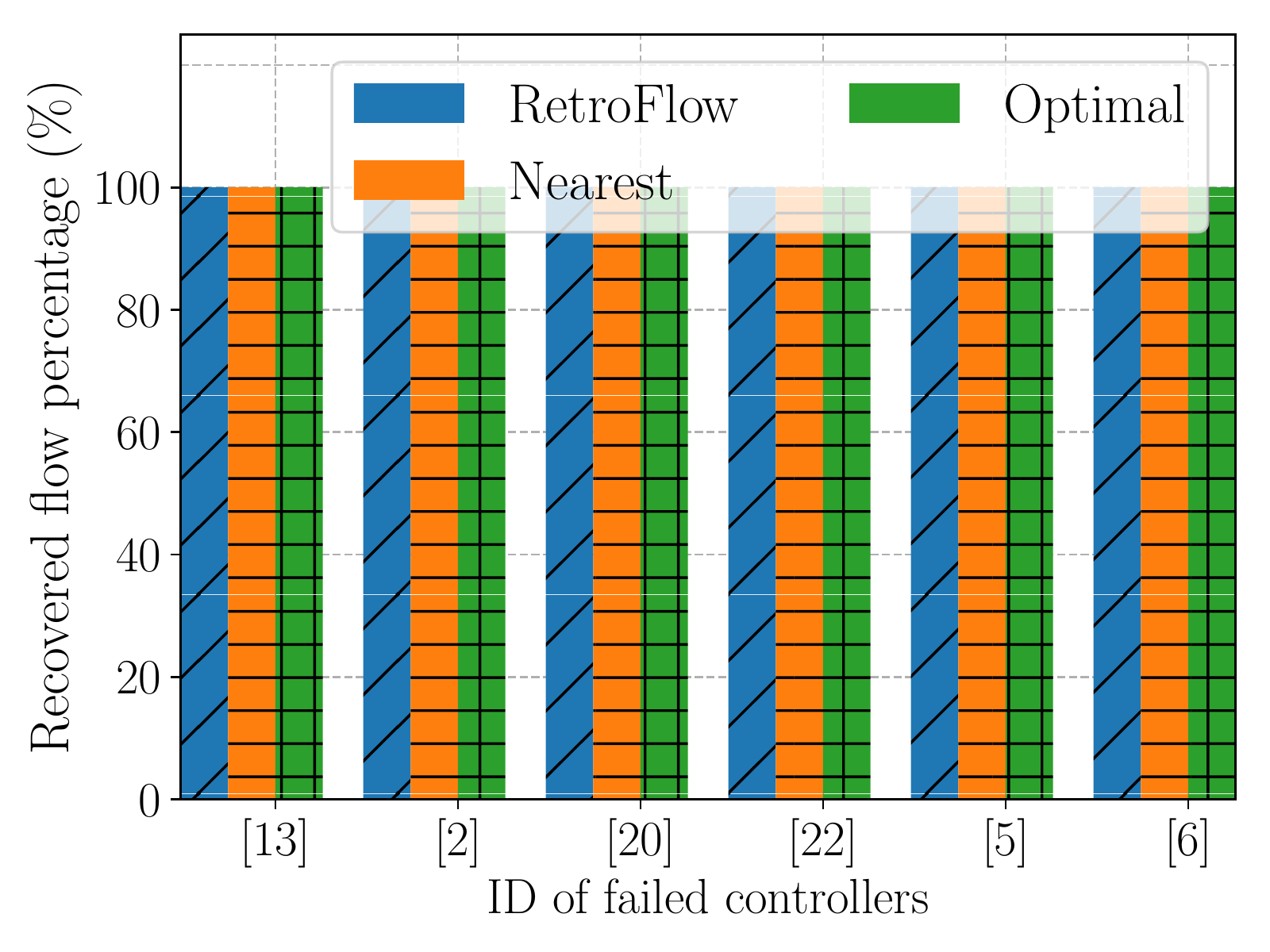}}
\hspace{0.05in}
\subfigure[Number of recovered offline switches. The lower, the better.]{
\includegraphics[width=\length]{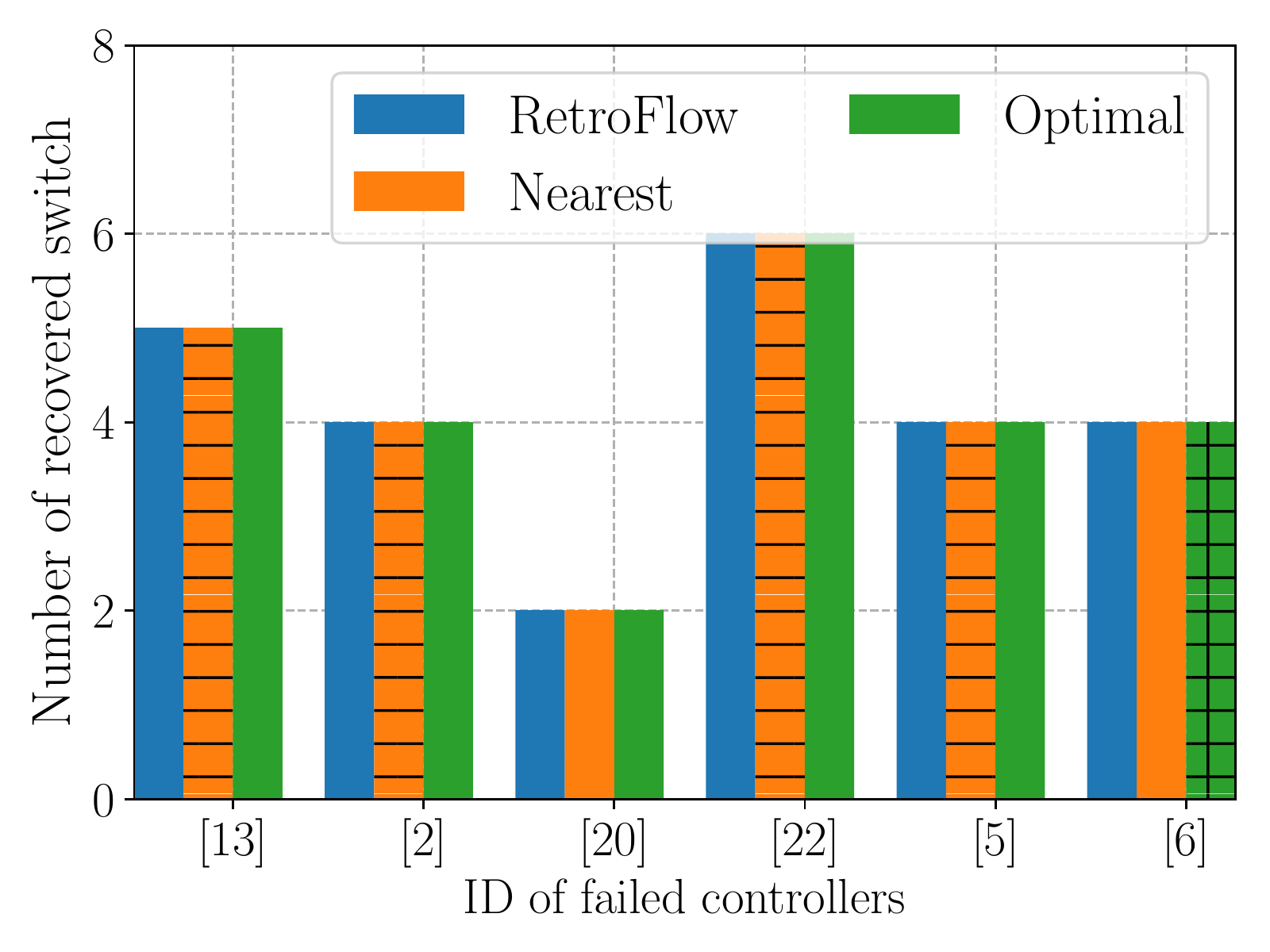}}
\hspace{0.05in}
\subfigure[Communication overhead. The lower, the better.]{
\includegraphics[width=\length]{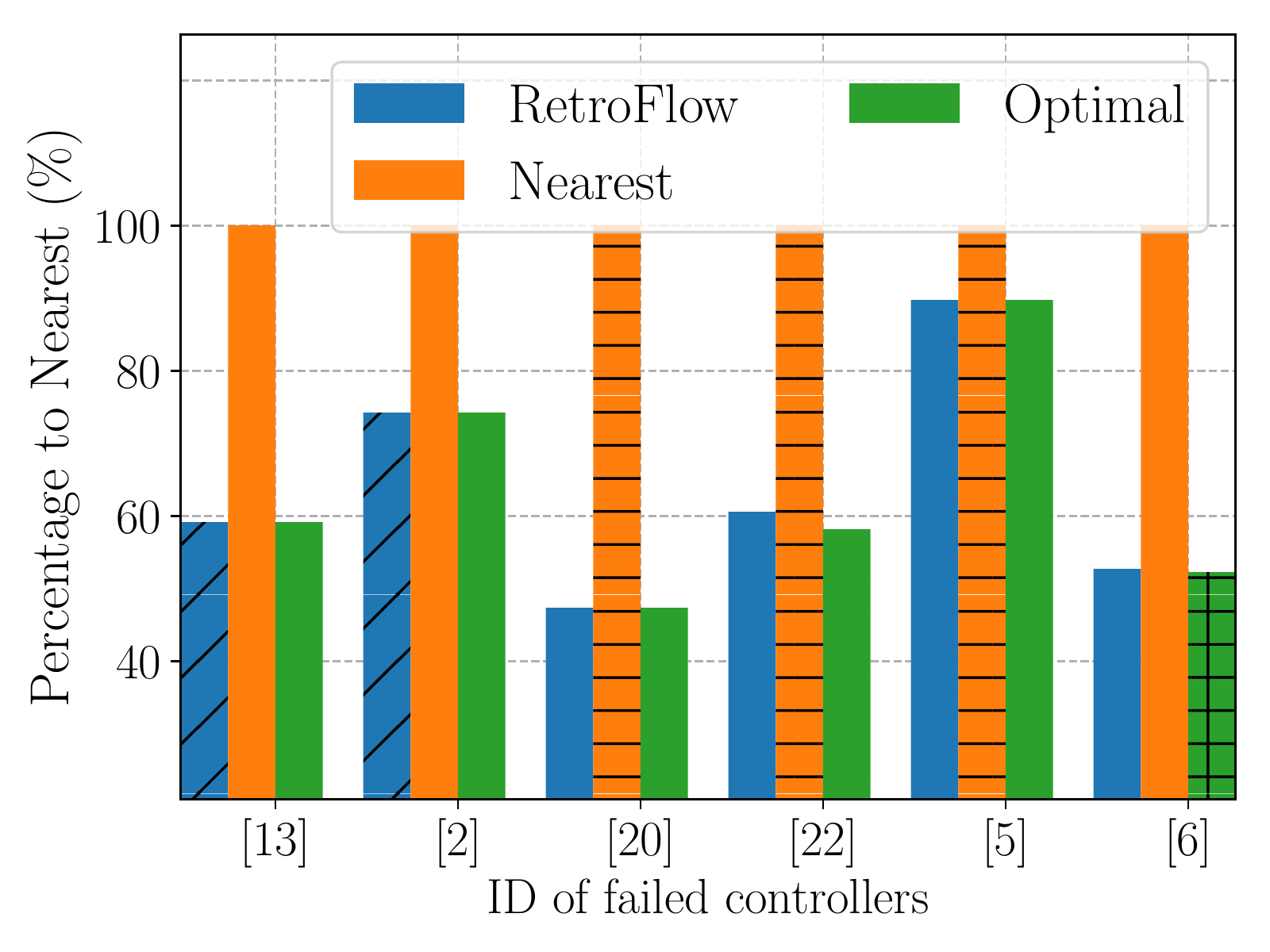}}
\hspace{0.05in}
\subfigure[Processing load of active controllers. The black dash line indicates the controller's processing ability.]{
\includegraphics[width=\length]{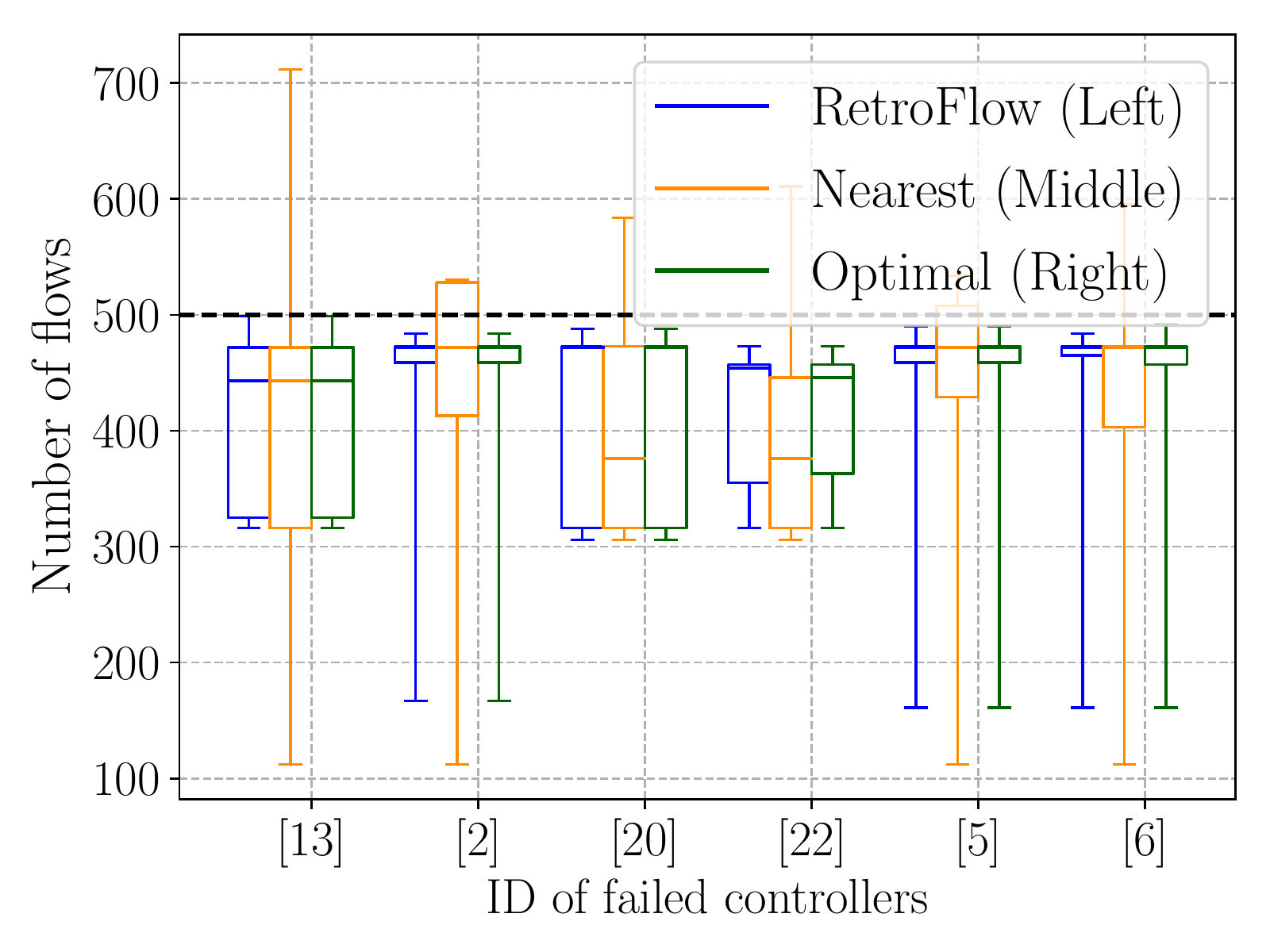}}
\vspace{-0.5cm}
\caption{Results of one controller fail and 100\% flows are set to recover. }
\label{fig:onecontroller}
\vspace{-0.4cm}
\end{figure*}

Details of \solution \ are summarized in Algorithm \ref{alg}, and Table \ref{table:notation} shows the notations used in the algorithm. In line 1, the sets $\mathcal{X}$, $\mathcal{Z}$, and $\mathcal{Y}$ are first set to be empty. In line 2, we start iteratively to find switches and their mappings. In line 3, for each iteration, we set $\delta$ to 0, and set $i_0$ and $j_{0}$ to NULL. In lines 5-14, we find the required switch. We first remove the existing programmable flows from each switch in $\mathcal{S}$ (lines 6-10) and find the required switch and update this switch's index to $i_0$ (lines 11-13). In lines 16-25, we test the mapping between switch $s_{i_0}$ and controller $C_{j}$. If controller $C_{j}$ has enough ability to control switch $s_{i_0}$, we select the controller as $C_{j_0}$, establish the mapping between switch $s_{i_0}$ and controller $C_{j_0}$, update the processing ability of controller $C_{j_0}$, and upgrade the set of programmable flows. In line 26, we remove the tested switch from the offline switches $\mathcal{S}$. In lines 27-29, if all switches are tested or the number of programmable flows reaches the flow programmability requirement, the algorithm jumps out of the iterations. In line 31, the result returns. 





\section{Simulation}
\label{sec:simulation}

\subsection{Simulation setup}  
We evaluate the performance of \solution \ with a real backbone topology named \topoa \ from Topology Zoo \cite{6027859}. The ATT topology is a national topology of US with 25 nodes and 112 links. In this topology, each node is given a unique ID with a latitude and a longitude. We calculate the distance between two nodes using Haversine formula \cite{robusto1957cosine} and use the distance divided by the propagation speed (i.e., 2$\times 10^8$ m/s) \cite{speed} to represent the propagation delay between the two nodes. In our simulation, each node is an SDN switch, and some selected nodes are further deployed controllers. Any two nodes have a traffic flow, and each flow is forwarded on its shortest path. We set the processing ability of a controller to 500. The default selection of controllers and default mapping between controllers and switches are obtained by solving an optimization problem, which aims to minimize the communication overhead among all switches and controllers. Table \ref{placement} shows the default relationship of controllers, switches, and 
the number of flows in the switches. 

\vspace{-.4cm}
\subsection{Comparison algorithms}
\begin{enumerate}
\item Optimal: it is the optimal solution of the OSCM problem that minimizes the communication overhead between offline switches and active controllers during controller failures. We solve the problem using GUROBI solver \cite{gurobi}.
\item Nearest: during controller failures, each offline switch maps to its nearest controller. This solution can minimize the propagation delay but could overload active controllers.

\item \solution: this algorithm is shown in Algorithm \ref{alg}. 
\end{enumerate}

\begin{figure}[t]
\centering
\subfigure[Percentage of programmable flows from offline switches. The higher, the better.]{
\includegraphics[width=3.5in]{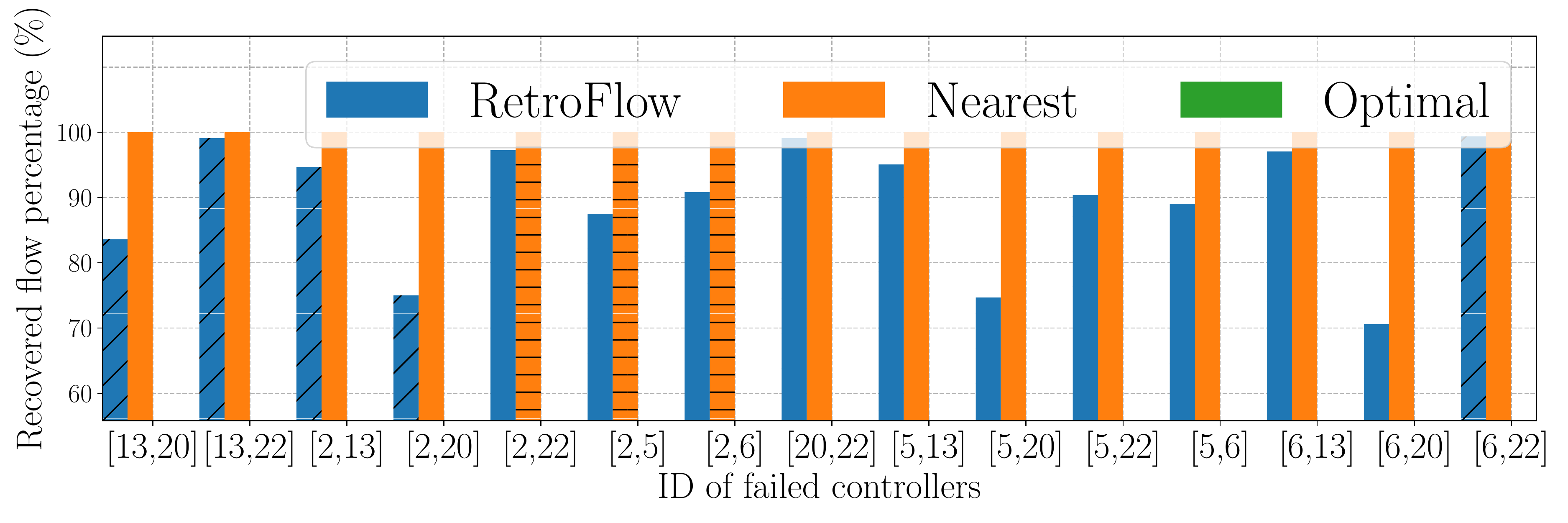}}
\subfigure[Number of recovered offline switches. The lower, the better.]{
\includegraphics[width=3.5in]{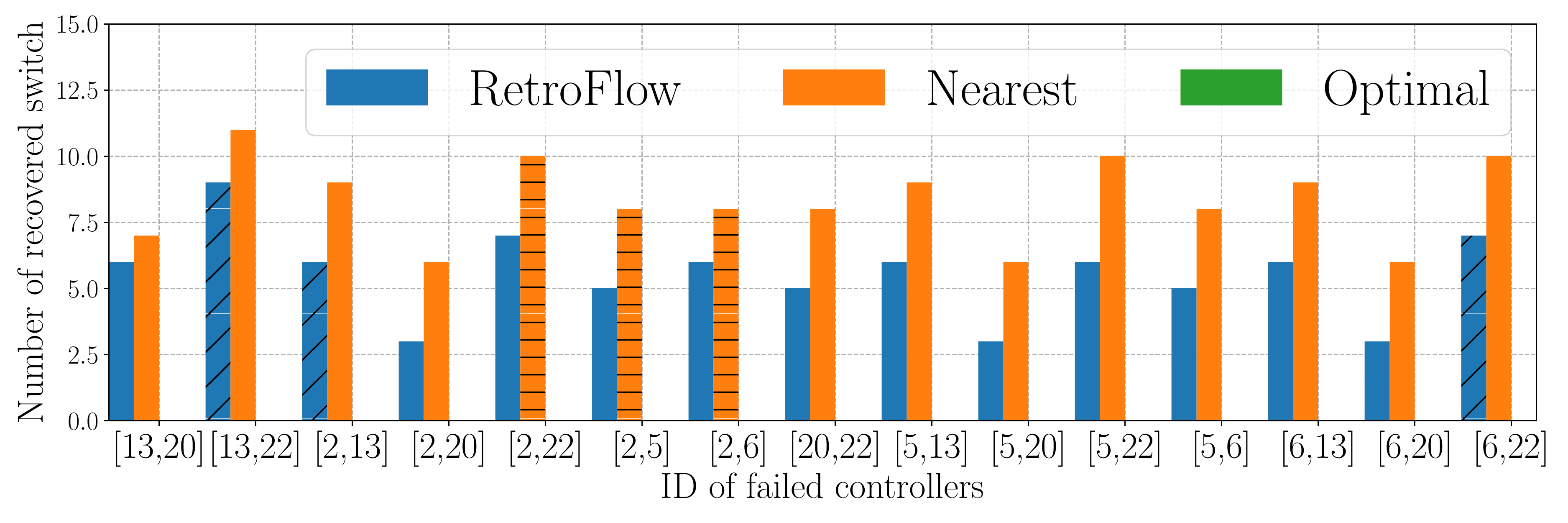}}
\subfigure[Communication overhead. The lower, the better.]{
\includegraphics[width=3.5in]{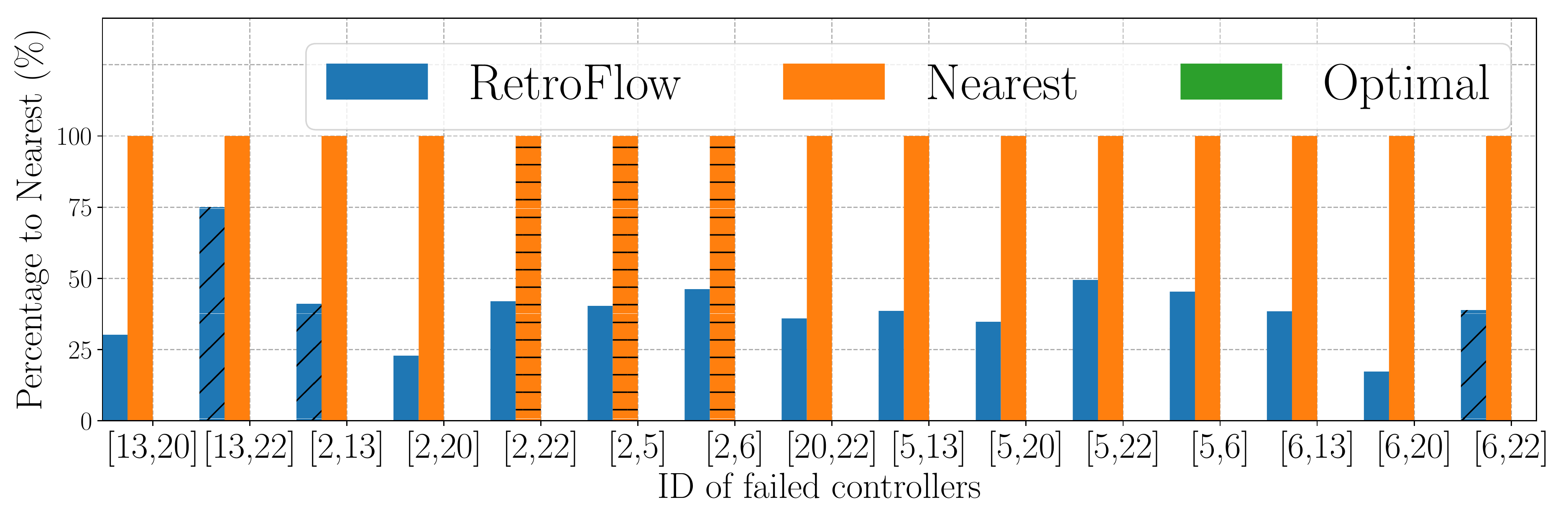}}
\subfigure[Processing load of active controllers. The black dash line indicates the controller's processing ability.]{
\includegraphics[width=3.5in]{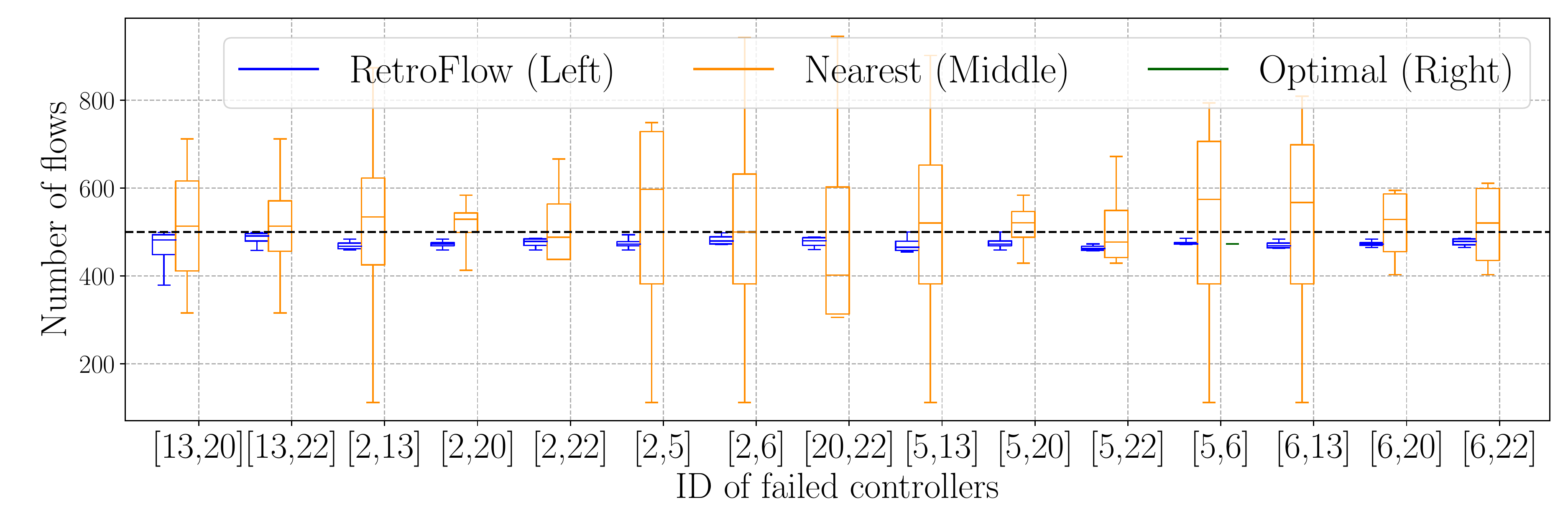}}
\vspace{-0.4cm}
\caption{Results of two controllers fail and 100\% flows are set to recover. Optimal cannot provide results.}
\label{fig:twocontroller100}
\vspace{-0.5cm}
\end{figure}

\vspace{-.5cm}
\subsection{Simulation results} 
In our simulation, the SDN control plane consists of six controllers. Many existing works consider only one controller failure \cite{hu2018adaptive}\cite{yao2013cascading}. \solution \ can cover a wide range of multiple controller failure scenarios. We compare \solution \ with other algorithms under two scenarios: (1) one controller failure and (2) two controllers failure. Scenario (1) is a moderate controller failure that active controllers have enough ability to handle all offline switches. Scenario (2) is a serious controller failure that active controllers are not able to handle all offline switches with their given processing abilities. Our performance metrics are the percentage of programmable flows from offline switches, communication overhead, and processing load of active controllers. We use Nearest as the baseline algorithm and normalize the metric of each algorithm to that of Nearest.

\subsubsection{One controller failure} 
Figure \ref{fig:onecontroller} shows the results of three algorithms when one of the six controllers fail. In Figures \ref{fig:onecontroller}(a) and (b), all three algorithms recover 100\% flows from offline switches, and they remap the same number of offline switches to active controllers. However, in Figure \ref{fig:onecontroller}(c), Optimal and \solution \ outperform Nearest in term of the communication overhead. This is because Optimal and \solution \ remap offline switches to their closest controllers with enough processing ability, while Nearest only considers the propagation delay to remap offline switches to controllers and thus could overload controllers, leading to long queueing delay for processing flow state pulling. The queueing delay setup follows the existing work \cite{xie2018cutting}. In Figure \ref{fig:onecontroller}(c), Optimal performs better than \solution \ because of its better switch-controller remapping. However, compared with Nearest, \solution \ can reduce the communication overhead up to 52.6\%. Figure \ref{fig:onecontroller}(d) shows the processing load of controllers. In this figure, Nearest experiences controller overload at all six cases. 

\begin{figure}[t]
\centering
\subfigure[Percentage of programmable flows from offline switches. The higher, the better.]{
\includegraphics[width=3.5in]{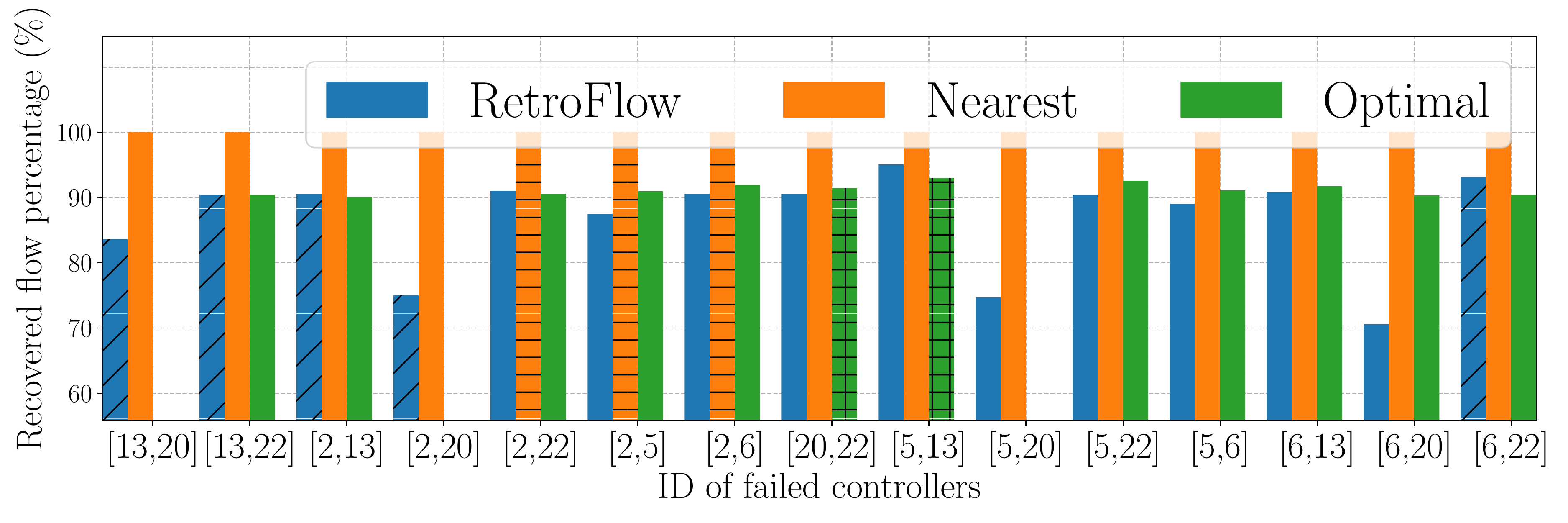}}
\subfigure[Number of recovered offline switches. The lower, the better.]{
\includegraphics[width=3.5in]{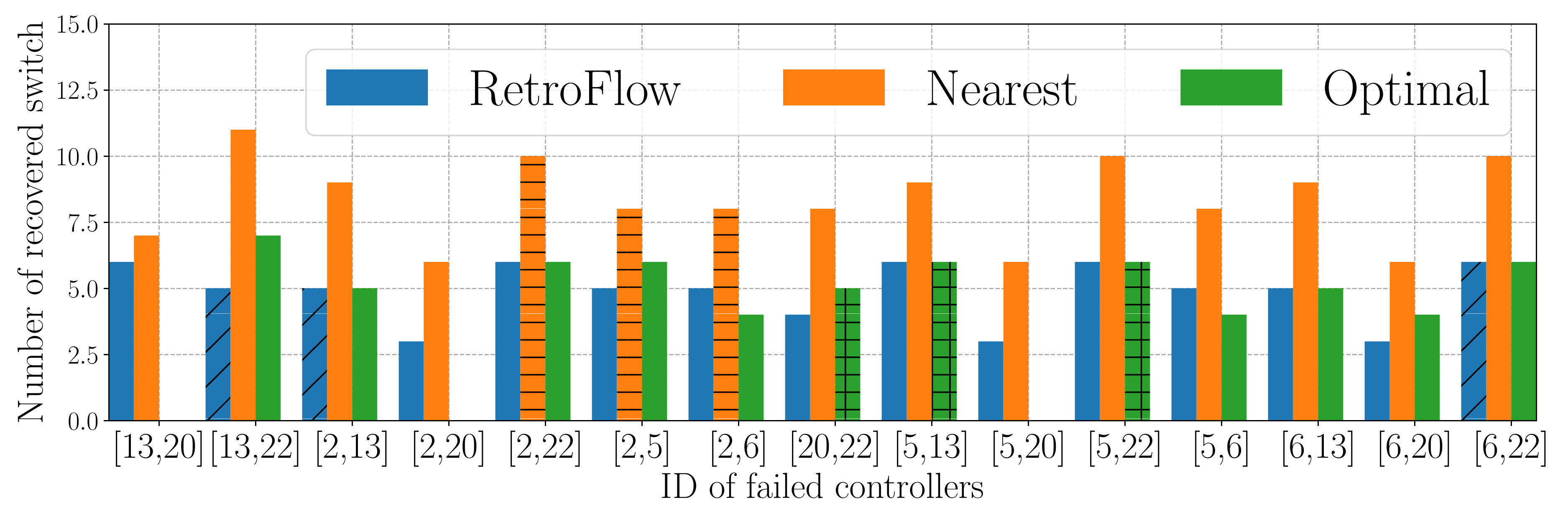}}
\subfigure[Communication overhead. The lower, the better.]{
\includegraphics[width=3.5in]{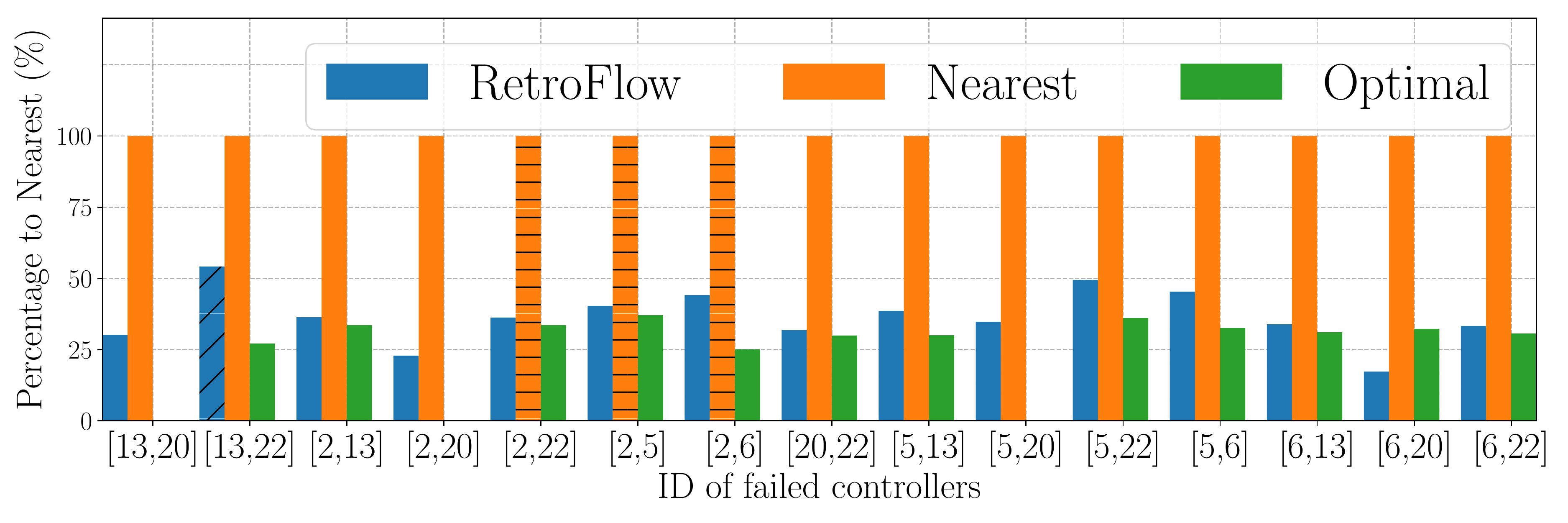}}
\subfigure[Processing load of active controllers. The black dash line indicates the controller's processing ability.]{
\includegraphics[width=3.43in]{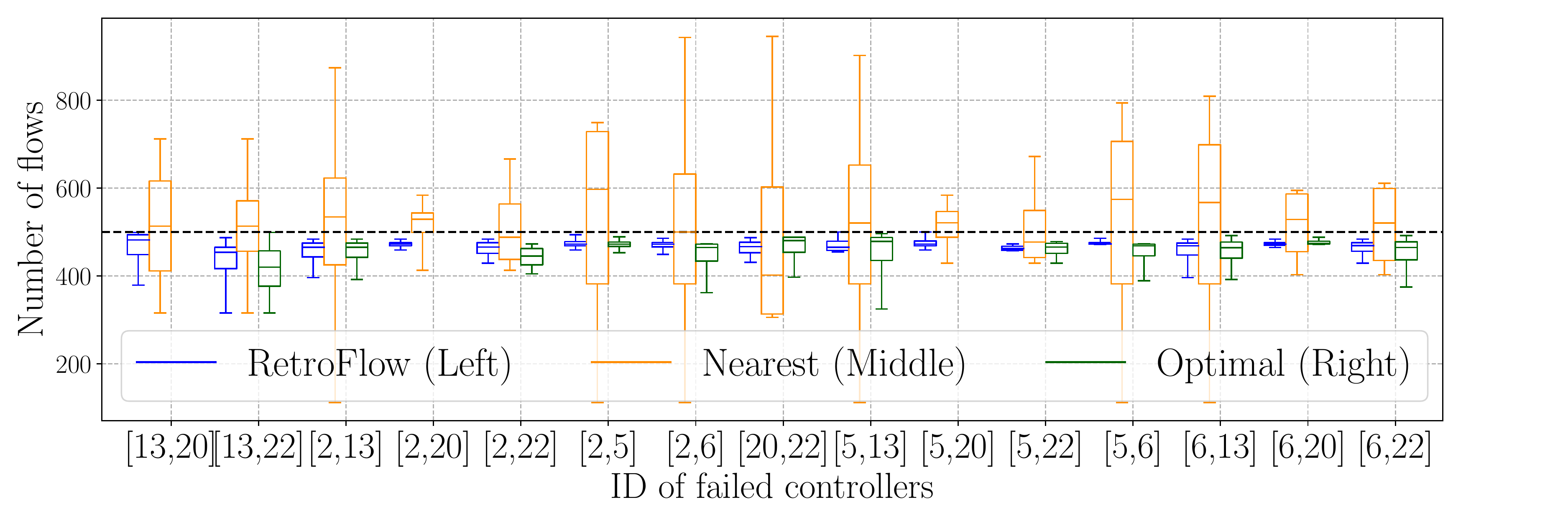}}
\vspace{-0.3cm}
\caption{Results when two controllers fail and 90\% flows are set to recover.}
\label{fig:twocontroller90}
\vspace{-0.6cm}
\end{figure}

\subsubsection{Two controllers failure} 
Figures \ref{fig:twocontroller100} and \ref{fig:twocontroller90} show the results of three algorithms when two of the six controllers fail. There are 15 combinations of the two controllers failure. In Figure \ref{fig:twocontroller100}, we require 100\% flows from offline switches to become programmable. Figure \ref{fig:twocontroller100}(a) shows the percentage of programmable flows from offline switches. In this figure, Optimal does not have results. Recall our problem has a constraint of not interrupting active controllers' normal operations. This constraint ensures each controller's processing load cannot exceed its processing ability, and under this constraint, Optimal cannot have a feasible solution even if all controllers reach their processing limits. Because \solution \ is a heuristic algorithm, it always has a solution. In this figure, \solution \ recovers flows in the range of 71 \%  to 99 \%. 

We analyze two representative failure cases: (1) the failure of controllers $C_{20}$ and $C_{22}$, and (2) the failure of controllers $C_{6}$ and $C_{20}$. In case (1), we have eight offline switches $s_{17}$-$s_{24}$. In Figure \ref{fig:twocontroller100}(b), \solution \ enables 99 \% flows to become programmable by recovering six offline switches, and Nearest recovers two more switches ($s_{18}$ and $s_{19}$) than \solution \ does but only controls 1\% more programmable flows. Because most of the flows in switches $s_{18}$ and $s_{19}$ have been recovered by remapping other six switches, remapping the two switches have only limited benefit. In case (2), we have six offline switches $s_0$, $s_1$, $s_6$, $s_7$, $s_{19}$, and $s_{20}$. In Figure \ref{fig:twocontroller100}(b), \solution \ recovers 71 \% flows by recovering three offline switches ($s_0$, $s_6$, and $s_7$), and Nearest recovers three more switches than \solution \ with an increase of 29\% more programmable flows. In this case, the left ability of controllers $C_2$, $C_5$, $C_{13}$, and $C_{22}$ are only 124, 184, 23, and 35 flows. Thus, under the controller's processing ability bound, \solution \ can only recover three switches, leading to a gap of programmable flow percentage between Nearest and \solution. Nearest enables 100\% programmable flows recovery at the cost of high communication overhead and controller overloading. In Figures \ref{fig:twocontroller100}(c) and (d), Nearest requires 25\% to 82\% more communication overhead due to the queueing delay of controller overloading.


If some controllers fail, it is unfair to overload other active controllers to take full responsibility for the failed controllers to control offline switches. Active controllers should only try their best to control offline switches. Based on this concern, in Figure \ref{fig:twocontroller90}, we require 90\% flows to become programmable. In this figure, Optimal has results for 12 of 15 cases. By reducing the number of programmable flows, the communication overhead of \solution \ reduces. Comparing the case of controllers $C_{13}$ and $C_{22}$ failure in Figures \ref{fig:twocontroller100}(c) and \ref{fig:twocontroller90}(c), \solution's overhead reduces from 74\% to 53\% because it maps five switches, which are three switches less than the scenario of 100\% flow recovery. When controllers $C_{20}$ and $C_{22}$ failure, \solution \ reduces the communication overhead up to 61.2\%.


%

\vspace{-0.3cm}
\section{Related Works} 
\label{sec:relatedworks} 
Pareto-based optimal controller-placement \cite{hock2013pareto_2} minimizes different objectives (e.g., the latency between switches and controllers, latency between controllers) under controller failures. Works in \cite{tanha2016enduring}\cite{perrot2016optimal} try to find the best trade-off between the performance and cost during the controller failure under several constraints (e.g., load balancing and QoS). The solution in \cite{alshamrani2018fault} proposes a controller placement model that ensures resiliency against the controller failure by minimizing the distance from a switch to its $i$-th closest controller. Capacitated Next Controller Placement \cite{killi2017capacitated} proposes a controller placement problem that not only considers the capacity and reliability of master controllers but also plans ahead for the master controller failure by considering a backup controller for each master controller. Different from all the aforementioned solutions, RetroFlow reduces the impact of controller failures by leveraging the features of hybrid SDN switches to maintain the advantage of SDN (i.e., flow programmability) and low communication overhead without overloading the rest active controllers.

%

\vspace{-0.3cm}
\section{Conclusion}
\label{sec:conclusion} 
In this paper, we propose \solution \ to jointly achieve resilient network control and flow programmablility during controller failures. \solution \ maintains active controllers' normal operations and programmablility of flows from offline switches while reducing the controllers' processing load from the offline switches by taking advantage of commercial hybrid SDN switches that support switches working under the legacy mode without controllers. By jointly considering the propagation delay and controllers' control cost in real time, \solution \ also achieves a low communication overhead among offline switches and active controllers. We hope that our work can inspire researchers to creatively utilize existing features in commercial SDN switches to better solve existing problems. 
\section{Acknowledgment}
This work was supported by the US NSF under Grants CNS-1618339, CNS-1617729, CNS-1814322, and CNS-1836772, the National Key Research and Development Program of China under Grant 2018YFB10- 03700, the NSFC under Grant 61836001, and the China Scholarship Council under Grants 201706370143 and 201806470060.
\bibliographystyle{ACM-Reference-Format}
\vspace{-.4cm}
\bibliography{retroflow}
\end{document}